%% file: main.tex
\pdfoutput=1
\documentclass[12pt,a4paper]{article}

\usepackage{ifthen} 
\newboolean{pdflatex}
\setboolean{pdflatex}{true} 

\newboolean{articletitles}
\setboolean{articletitles}{true} 

\newboolean{uprightparticles}
\setboolean{uprightparticles}{false} 

\newcommand{\aunit}[1]{\ensuremath{\text{\,#1}}}       
\newcommand{\tev}{\aunit{Te\kern -0.1em V}\xspace}
\newcommand{\gev}{\aunit{Ge\kern -0.1em V}\xspace}
\newcommand{\gevc}{\ensuremath{\aunit{Ge\kern -0.1em V\!/}c}\xspace}
\newcommand{\ie}{\mbox{\itshape i.e.}\xspace}

\def\lhcb   {\mbox{LHCb}\xspace}

\def\alice  {\mbox{ALICE}\xspace}
\def\thebaroffset{0.0em}
\newcommand{\offsetoverline}[2][\thebaroffset]{\kern #1\overline{\kern -#1 #2}}%
\def\sPlot{\mbox{\em sPlot}\xspace}

\def\PD      {\ensuremath{D}\xspace}                 
\def\Pmu         {\ensuremath{\mu}\xspace}                 
\def\PB      {\ensuremath{B}\xspace}                 
\def\PJ      {\ensuremath{J}\xspace}                 
\def\Pb      {\ensuremath{b}\xspace}                 
\def\Pp      {\ensuremath{p}\xspace}                 
\def\Pc      {\ensuremath{c}\xspace}                 
\def\Ps      {\ensuremath{s}\xspace}                 
\def\PK      {\ensuremath{K}\xspace}                 
\def\Ppi     {\ensuremath{\pi}\xspace}                 
\def\Ppsi    {\ensuremath{\psi}\xspace}                 
\def\PLambda {\ensuremath{\Lambda}\xspace}                 
\mathchardef\PLambda="7103
\mathchardef\PUpsilon="7107

\def\Dbar    {{\ensuremath{\offsetoverline{\PD}}}\xspace}
\def\B       {{\ensuremath{\PB}}\xspace}
\def\D       {{\ensuremath{\PD}}\xspace}
\def\cquark    {{\ensuremath{\Pc}}\xspace}
\def\bquark    {{\ensuremath{\Pb}}\xspace}
\def\Dz      {{\ensuremath{\D^0}}\xspace}
\def\squark    {{\ensuremath{\Ps}}\xspace}
\def\Lz          {{\ensuremath{\PLambda}}\xspace}
\def\Lc          {{\ensuremath{\Lz^+_\cquark}}\xspace}
\def\Dzb     {{\ensuremath{\Dbar{}^0}}\xspace}
\def\jpsi     {{\ensuremath{{\PJ\mskip -3mu/\mskip -2mu\Ppsi}}}\xspace}
\def\PUpsilon    {\ensuremath{\Upsilon}\xspace}                 
\def\Upsilonres  {{\ensuremath{\PUpsilon}}\xspace}
\def\Bz      {{\ensuremath{\B^0}}\xspace}
\def\Bp      {{\ensuremath{\B^+}}\xspace}
\def\Lb           {{\ensuremath{\Lz^0_\bquark}}\xspace}
\def\kaon    {{\ensuremath{\PK}}\xspace}
\def\Kp      {{\ensuremath{\kaon^+}}\xspace}
\def\Km      {{\ensuremath{\kaon^-}}\xspace}
\def\Kpm     {{\ensuremath{\kaon^\pm}}\xspace}
\def\Kmp     {{\ensuremath{\kaon^\mp}}\xspace}
\def\KS      {{\ensuremath{\kaon^0_{\mathrm{S}}}}\xspace}
\def\pion   {{\ensuremath{\Ppi}}\xspace}

\def\pip    {{\ensuremath{\pion^+}}\xspace}
\def\pim    {{\ensuremath{\pion^-}}\xspace}
\def\pipm   {{\ensuremath{\pion^\pm}}\xspace}
\def\pimp   {{\ensuremath{\pion^\mp}}\xspace}
\def\mup        {{\ensuremath{\Pmu^+}}\xspace}
\def\mun        {{\ensuremath{\Pmu^-}}\xspace} 
\def\Dstar   {{\ensuremath{\D^*}}\xspace}
\def\Dstarp  {{\ensuremath{\D^{*+}}}\xspace}
\def\Ds      {{\ensuremath{\D^+_\squark}}\xspace}
\def\Dp      {{\ensuremath{\D^+}}\xspace}
\def\pt         {\ensuremath{p_{\mathrm{T}}}\xspace}
\def\sqs   {\ensuremath{\protect\sqrt{s}}\xspace}
\def\sqsnn {\ensuremath{\protect\sqrt{s_{\scriptscriptstyle\text{NN}}}}\xspace}
\def\logipchisq{\ensuremath{\log_{10}\left(\chi^2_{\text{IP}}\right)}}
\def\proton      {{\ensuremath{\Pp}}\xspace}
\def\nb {\aunit{nb}\xspace}
\def\invnb {\ensuremath{\nb^{-1}}\xspace}
\def\pythia     {\mbox{\textsc{Pythia}}\xspace}

\def\evtgen     {\mbox{\textsc{EvtGen}}\xspace}
\def\photos     {\mbox{\textsc{Photos}}\xspace}
\def\geant      {\mbox{\textsc{Geant4}}\xspace}
\def\deriv {\ensuremath{\mathrm{d}}}
\newcommand{\decay}[2]{\ensuremath{#1\!\to #2}\xspace} 
\def\dkpicf     {\decay{\Dz}{\Km\pip}}
\def\mbarn{\aunit{mb}\xspace}
\newcommand{\vs}{\mbox{\itshape vs.}\xspace}
\newcommand{\lhcborcid}[1]{\href{https://orcid.org/#1}{\hspace*{0.1em}\raisebox{-0.45ex}{\includegraphics[width=1em]{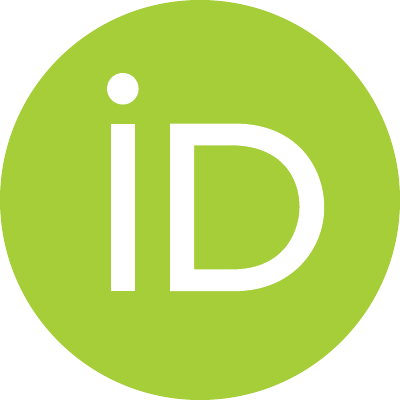}}}}

\def\paperauthors{LHCb collaboration} 
\def\paperasciititle{Measurement of prompt D0 nulcear modification factor in pPb collisions at sNN=8.16TeV} 
\def\papertitle{Measurement of the prompt \Dz nuclear modification factor in $p$Pb collisions at $\sqsnn=8.16\tev$} 
\def\paperkeywords{{High Energy Physics}, {LHCb}} 
\def\papercopyright{\the\year\ CERN for the benefit of the LHCb collaboration} 
\def\paperlicence{CC BY 4.0 licence}
\def\paperlicenceurl{https://creativecommons.org/licenses/by/4.0/}

\usepackage[top=1in, bottom=1.25in, left=1in, right=1in]{geometry}
\usepackage{microtype}
\usepackage{lineno}  
\usepackage{xspace} 
\usepackage{caption} 

\usepackage{graphicx}  
\usepackage{color}
\usepackage{colortbl}

\usepackage{amsmath} 
\usepackage{amssymb}
\usepackage{amsfonts}
\usepackage{upgreek} 

\newcommand*\patchAmsMathEnvironmentForLineno[1]{%
    \expandafter\let\csname old#1\expandafter\endcsname\csname #1\endcsname
    \expandafter\let\csname oldend#1\expandafter\endcsname\csname
    end#1\endcsname
    \renewenvironment{#1}%
    {\linenomath\csname old#1\endcsname}%
    {\csname oldend#1\endcsname\endlinenomath}%
    }
    \newcommand*\patchBothAmsMathEnvironmentsForLineno[1]{%
        \patchAmsMathEnvironmentForLineno{#1}%
        \patchAmsMathEnvironmentForLineno{#1*}%
        }
        \AtBeginDocument{%
            \patchBothAmsMathEnvironmentsForLineno{equation}%
            \patchBothAmsMathEnvironmentsForLineno{align}%
            \patchBothAmsMathEnvironmentsForLineno{flalign}%
            \patchBothAmsMathEnvironmentsForLineno{alignat}%
            \patchBothAmsMathEnvironmentsForLineno{gather}%
            \patchBothAmsMathEnvironmentsForLineno{multline}%
            \patchBothAmsMathEnvironmentsForLineno{eqnarray}%
            }

            \usepackage{hyperxmp}

            \usepackage[pdftex,
            pdfauthor={\paperauthors},
            pdftitle={\paperasciititle},
            pdfkeywords={\paperkeywords},
            pdfcopyright={Copyright (C) \papercopyright},
            pdflicenseurl={\paperlicenceurl}]{hyperref}

            \usepackage[colorinlistoftodos,textsize=scriptsize]{todonotes}

            \usepackage[bottom,flushmargin,hang,multiple]{footmisc}

            \usepackage[all]{hypcap} 

            \usepackage{cite} 
            \usepackage{mciteplus}
            \usepackage[top=1in, bottom=1.25in, left=1in, right=1in]{geometry}
            \usepackage{longtable} 
            \usepackage{rotating}
            \usepackage{arydshln}
            \usepackage{rotating}
            \usepackage{lineno}
            
\begin{document}

            \renewcommand{\thefootnote}{\fnsymbol{footnote}}
            \setcounter{footnote}{1}

            \begin{titlepage}
                \pagenumbering{roman}

                \vspace*{-1.5cm}
                \centerline{\large EUROPEAN ORGANIZATION FOR NUCLEAR RESEARCH (CERN)}
                \vspace*{1.5cm}
                \noindent
                \begin{tabular*}{\linewidth}{lc@{\extracolsep{\fill}}r@{\extracolsep{0pt}}}
                    \ifthenelse{\boolean{pdflatex}}
                    {\vspace*{-1.5cm}\mbox{\!\!\!\includegraphics[width=.14\textwidth]{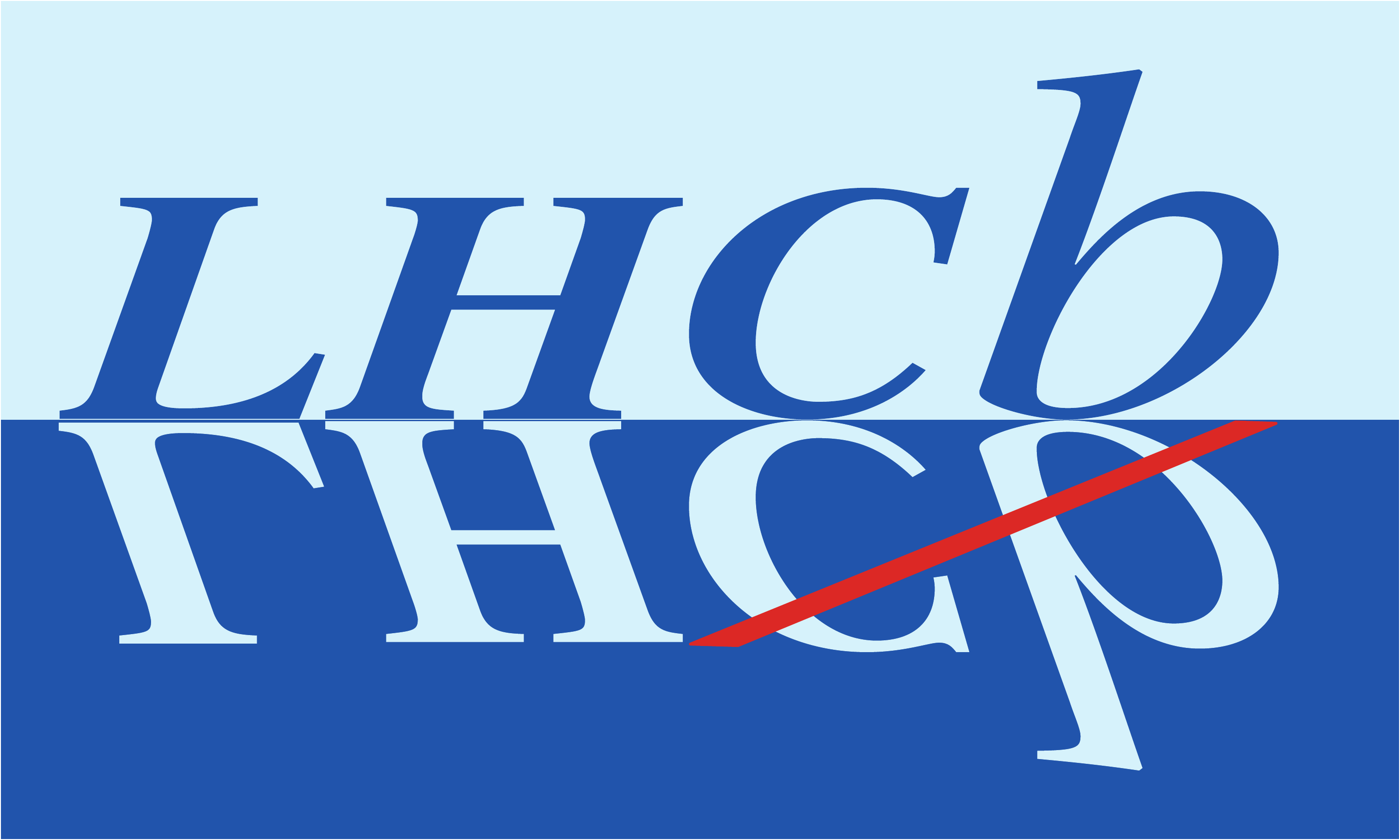}} & &}%
                    {\vspace*{-1.2cm}\mbox{\!\!\!\includegraphics[width=.12\textwidth]{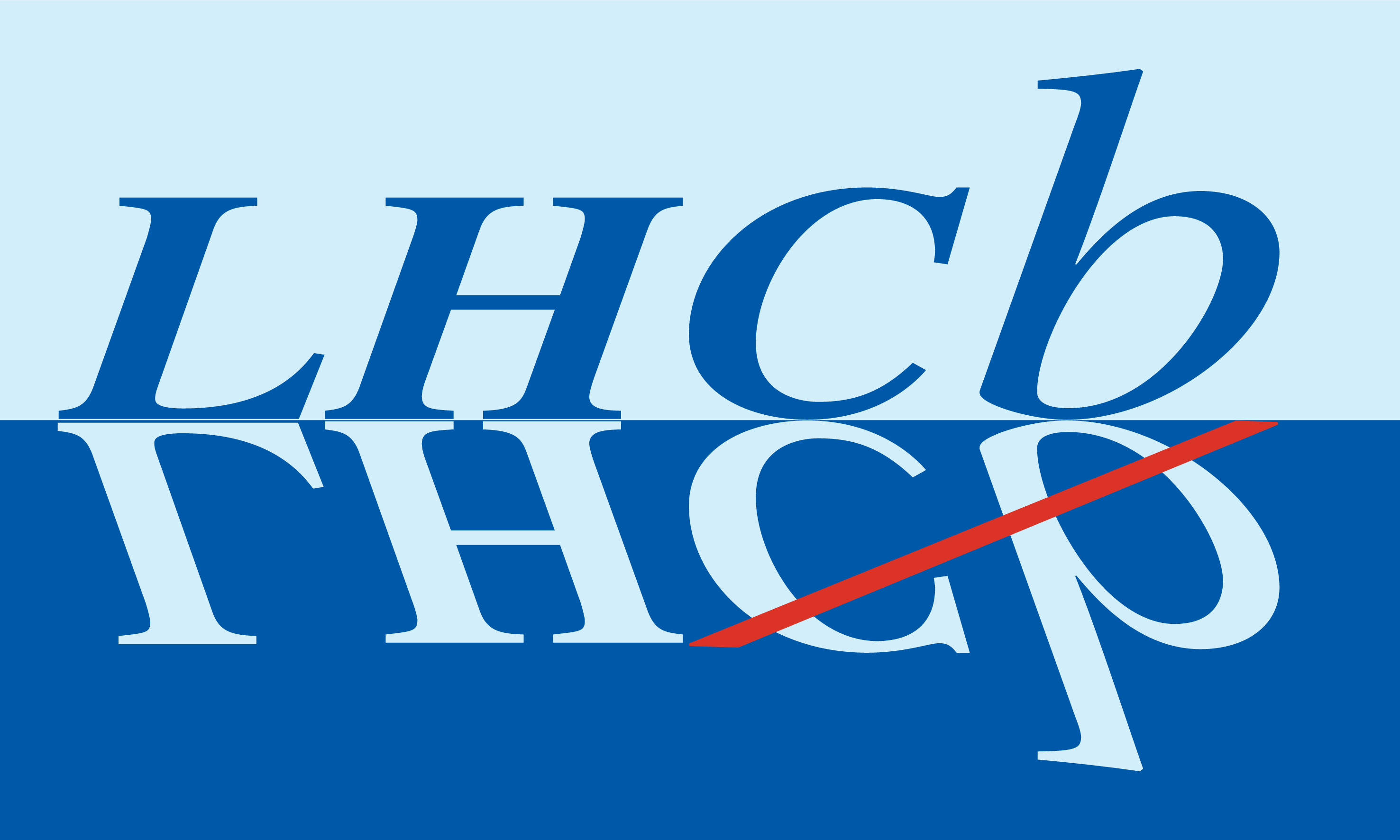}} & &}%
                    \\
                    & & CERN-EP-2022-082 \\  
                    & & LHCb-PAPER-2022-007 \\  
                    & & \today \\ 
                    & & \\
                \end{tabular*}

                \vspace*{4.0cm}

                {\normalfont\bfseries\boldmath\huge
                \begin{center}
                    \papertitle 
                \end{center}
                }

                \vspace*{2.0cm}

                \begin{center}
                    \paperauthors\footnote{Authors are listed at the end of this Letter.}
                \end{center}

                \vspace{\fill}

                \begin{abstract}
                    \noindent
                    The production of prompt \Dz mesons in proton-lead
                    collisions in both the forward and backward rapidity regions
                    at a center-of-mass energy per nucleon pair of
                    \mbox{$\sqsnn = 8.16\tev$} is measured by the \lhcb experiment.
                    The nuclear modification factor of prompt \Dz mesons 
                    is determined as a function of the transverse momentum \pt,
                    and the rapidity in the nucleon-nucleon center-of-mass frame $y^*$.
                    In the forward rapidity region, significantly suppressed production
                    with respect to $pp$ collisions is measured,
                    which provides significant constraints on models of nuclear parton distributions and hadron production 
                    down to the very low Bjorken-$x$ region of $\sim 10^{-5}$.
                    In the backward rapidity region,
                    a suppression with a significance of $2.0-3.8$ standard deviations
                    compared to nPDF expectations is found
                    in the kinematic region of $\pt>6\gevc$ and $-3.25<y^*<-2.5$,
                    corresponding to $x\sim 0.01$.
                \end{abstract}

                \vspace*{2.0cm}

                \begin{center}
                    Published in
                    Phys.~Rev.~Lett. {\bf 131} (2023) 102301.
                \end{center}

                \vspace{\fill}

                {\footnotesize 
                \centerline{\copyright~\papercopyright. \href{\paperlicenceurl}{\paperlicence}.}}
                \vspace*{2mm}

            \end{titlepage}


            \newpage
            \setcounter{page}{2}
            \mbox{~}

            \renewcommand{\thefootnote}{\arabic{footnote}}
            \setcounter{footnote}{0}

            \cleardoublepage


            \pagestyle{plain} 
            \setcounter{page}{1}
            \pagenumbering{arabic}


            Charm and beauty quarks are produced in the early stage of ultra-relativistic
            heavy-ion collisions
            and are strongly affected by the presence of deconfined hot nuclear matter,
            known as quark-gluon plasma (QGP)~\cite{Yagi:2005yb},
            as well as by cold nuclear matter (CNM) effects.
            The latter can be studied in proton-nucleus collisions
            where QGP effects are not expected to be dominant.
            Heavy-flavor hadrons, \ie hadrons containing one or more heavy quark,
            are affected by CNM effects at all stages of their production.
            At LHC energies, the most relevant effect is from the initial state,
            where 
            the parton distribution functions in a nuclear environment (nPDF)~\cite{Eskola:2009uj,deFlorian:2003qf,Hirai:2007sx}
            differ from those in isolated nucleons
            at all values of Bjorken momentum fraction $x$. 
            Parton density decreases at $x\lesssim 0.1$ due to nuclear shadowing, and increases at $0.1\lesssim x\lesssim 0.3$ as a result of antishadowing~\cite{Armesto:2006ph}.
            Therefore different effects are expected to be relevant at different intervals of rapidity, which is strongly correlated with $x$.
            The parton distributions at small $x$ can also be described
            by the color-glass condensate effective theory (CGC) as a saturated gluonic system~\cite{Gelis:2010nm}.
            Moreover, multiple scattering and energy loss may occur
            when the incoming partons and the 
            heavy quarks traverse the nuclear medium~\cite{Vitev:2007ve,Kang:2014hha,Arleo:2021bpv}.
            Other initial-state or even final-state effects~\cite{Zhang:2019dth,Zhao:2020wcd}
            may also modify the kinematic distributions of produced heavy-flavor hadrons,
            as suggested by their surprisingly large spatial anisotropy in momentum
            in high-multiplicity $p$Pb collisions~\cite{Sirunyan:2018toe, Sirunyan:2018kiz}.

            The LHCb collaboration has recently measured the production cross-section
            of various heavy-flavor hadrons in $p$Pb collisions at forward rapidity,
            including the production of prompt \Dz and \Lc hadrons at
            a center-of-mass energy per nucleon pair of \mbox{$\sqsnn=5.02$\tev}~\cite{LHCb-PAPER-2017-015,LHCb-PAPER-2018-021},
            and the production of \jpsi, \Bz, \Bp, \Lb and $\Upsilonres(nS)$ states
            at \mbox{$\sqsnn=8.16$\tev}~\cite{LHCb-PAPER-2017-014,LHCb-PAPER-2018-048,LHCb-PAPER-2018-035}. The ALICE collaboration has measured open charm production at midrapidity
            at \mbox{$\sqsnn=5.02\tev$}~\cite{ALICE:2014xjz,ALICE:2016cpm,ALICE:2016yta,ALICE:2019fhe,ALICE:2020wfu,ALICE:2020wla}. 
            Other measurements of heavy-flavor production in $p$Pb collisions at the LHC
            are also reported~\cite{ALICE:2017fsl,ALICE:2020vjy,ALICE:2019qie,ALICE:2018mml,ALICE:2018szk,ALICE:2014ict,ATLAS:2017prf,CMS:2022wfi,CMS:2017exb,CMS:2015sfx,CMS:2016wma,CMS:2015gcq}.
            CNM effects have also been investigated with heavy-quark production at the RHIC collider in $d$Au collisions at \mbox{\sqsnn = 200\gev}~\cite{STAR:2004ocv,PHENIX:2012hww}.
            These measurements have led to significantly reduced uncertainties of nPDFs
            in the small-$x$ region~\cite{Eskola:2019bgf,Khalek:2022zqe},
            especially with the constraints from the LHCb $\Dz$ measurements at \mbox{$\sqsnn=5.02\tev$}~\cite{LHCb-PAPER-2017-015}.

            This Letter reports the measurement of the production cross-section
            and the nuclear modification factor $R_{p\mathrm{Pb}}$
            of prompt \Dz mesons in $p$Pb collisions at $\sqsnn=8.16$\tev performed
            with the \lhcb detector~\cite{Alves:2008zz}.
            The quantity $R_{p\mathrm{Pb}}$ is defined as the ratio of the cross-section in $p$Pb collisions
            to the corresponding cross-section in $pp$ collisions
            scaled by the mass number of Pb.
            Prompt mesons are those directly produced in proton-lead collisions or from strong decays of excited charm hadrons,
            rather than from decays of beauty hadrons.
            This measurement uses a data sample 20 times larger than
            that used for the LHCb \Dz measurements at $\sqsnn=5.02\tev$\cite{LHCb-PAPER-2017-015}.
            The results can be incorporated into global fits together with all other relevant measurements to improve nPDF parameterizations and to test other possible CNM effects.

            The \lhcb detector is a single-arm forward
            spectrometer designed for studying heavy-flavor particles,
            described in detail in Refs.~\cite{Alves:2008zz,LHCb-DP-2014-002}.
            The data sample for this analysis consists of \ensuremath{p\mathrm{Pb}} collisions
            collected with the \lhcb detector at the end of 2016, including two different configurations:
            forward collisions (\proton beam coming from upstream of the vertex detector)
            and backward collisions (\proton beam coming from downstream of the vertex detector),
            corresponding to an integrated luminosity of $12.2 \pm 0.3$\invnb ($18.6 \pm 0.5$\invnb)
            for forward (backward) collisions~\cite{LHCb-PAPER-2018-048,LHCb-PAPER-2014-047}.
            The forward (backward) configuration data cover a positive (negative) rapidity range of $1.5 < y^* < 4.0$ ($-5.0 < y^* < -2.5$), corresponding to an $x$ coverage of approximately $ 10^{-5}-10^{-3}$ ($10^{-2}-10^{-1}$) 
            for the partons of the Pb nucleus,
            with the positive $z$-axis defined as the direction of the proton beam.

            Simulation samples are required to model the effects of the detector acceptance
            and the selection requirements.
            The \Dz mesons are generated using
            \pythia8~\cite{Sjostrand:2006za,*Sjostrand:2007gs} with a specific \lhcb configuration~\cite{LHCb-PROC-2010-056} and embedded into minimum-bias $p$Pb events
            from the \mbox{\textsc{EPOS-LHC}}
            generator\cite{PhysRevC.92.034906}.
            Decays of unstable particles
            are described by \evtgen~\cite{Lange:2001uf}, in which final-state
            radiation is generated using \photos~\cite{Golonka:2005pn}.
            The interaction of the generated particles with the detector,
            and its response, are implemented using the \geant
            toolkit~\cite{Allison:2006ve, *Agostinelli:2002hh} as described in
            Ref.~\cite{LHCb-PROC-2011-006}.

            The double differential cross-section for prompt \Dz production is measured as a function of $y^*$,
            the rapidity in the nucleon-nucleon center-of-mass frame,
            and \pt, the transverse momentum with respect to the beam direction.
            The quantity $y^*$ is related to the rapidity in the laboratory frame
            $y_\mathrm{lab}$ by $y^* = y_\mathrm{lab}-0.465$ for $p$Pb collisions.
            The differential cross-section in a given (\pt, $y^*$) interval is defined as
            \begin{equation}\label{eqn:cross-def}
                \frac{\deriv^2 \sigma}{\deriv \pt \deriv y^*}\equiv
                \frac{N(\decay{\Dz}{\Kmp\pipm})+N(\decay{\Dzb}{\Kpm\pimp})}
                {\mathcal{L}\times\varepsilon_{\text{tot}}\times\mathcal{B}(\decay{\Dz}{\Kmp\pipm})\times\Delta\pt \times \Delta y^*}~,
            \end{equation}
            where $N(\decay{\Dz}{\Kmp\pipm})$ and $N(\decay{\Dzb}{\Kpm\pimp})$ are the \Dz and \Dzb signal yields,
            $\mathcal{L}$ is the integrated luminosity,
            $\varepsilon_{\mathrm{tot}}$ is the total efficiency,
            \mbox{$\mathcal{B}(\decay{\Dz}{\Kmp\pipm})=(3.96\pm0.03)\%$}
            is the sum of branching fractions for the decays \dkpicf and \decay{\Dz}{\Kp\pim}~\cite{PDG2022}, and
            $\Delta \pt$ and $\Delta y^*$ are the \pt and $y^*$ interval widths.
            The \Dz mesons are reconstructed
            through the \decay{\Dz}{\Km\pip} and 
            the doubly Cabibbo-suppressed \decay{\Dz}{\Kp\pim} decay channels and their charge conjugates
            \footnote{The branching fraction of $\decay{\Dz}{\Kp\pim}$ channel
            is two orders of magnitude smaller than that of $\decay{\Dz}{\Km\pip}$.}.
            The measurement is performed within a \pt range of $0<\pt<30\gevc$, and the rapidity range defined above.
            Throughout the analysis,
            the measurements are for the combined sample of \Dz and \Dzb mesons.
            The signal yields and the total efficiency
            are determined in each kinematic interval.

            The \Dz candidates are built from \Kmp and \pipm candidate tracks.
            The selection criteria are similar to those used in \Dz production measurements
            in $p\mathrm{Pb}$ collisions at \mbox{$\sqsnn = 5.02\tev$}~\cite{LHCb-PAPER-2017-015}.
            The reconstructed \Kmp and \pipm tracks
            are required to have transverse momentum greater than $0.4\gevc$.
            Both tracks are also required to be of good quality,
            come from a common vertex,
            and pass particle identification (PID) requirements.

            The inclusive \Dz signal yield is the sum of the prompt \Dz mesons
            and those produced in the decays of \bquark hadrons,
            denoted ``{\it from-\bquark}''.
            This inclusive yield is determined using
            an extended unbinned maximum-likelihood fit to
            the distribution of the \kaon\pion invariant mass, $M(\kaon \pion)$.
            The $M(\kaon \pion)$ distribution of the signal is described by a sum
            of a Crystal Ball function~\cite{Skwarnicki:1986xj} and a Gaussian function
            sharing a common mean value,
            while the background shape is described by a linear function,
            following the measurement of Ref.~\cite{LHCb-PAPER-2017-015}.
            The prompt signal yield is determined
            by fitting the distribution of  \logipchisq\ of the \Dz candidates,
            where $\chi^2_{\mathrm{IP}}$ is defined as the difference in the vertex-fit $\chi^2$
            of a given primary vertex reconstructed with and without the \Dz candidate under consideration. 
            The background component in the \logipchisq\ distribution
            is subtracted using the \sPlot technique~\cite{pivk2005plots} with $M(\kaon \pion)$ as the discriminating variable.
            The shapes of the \logipchisq\ distribution
            corresponding to the prompt and {\it from-\bquark} components
            are described independently by Bukin functions~\cite{bukin2007fitting},
            which are asymmetric functions with tails described by Gaussian functions.
            The parameters of the functions describing the prompt
            and {\it from-\bquark} components are fixed to those from simulation.
            The invariant-mass and \logipchisq\ distributions of the forward and backward samples
            are given in the Supplemental Material~\cite{ref:SuppMat}.

            The total efficiency $\varepsilon_{\mathrm{tot}}$ is the product of
            the geometrical acceptance of the detector,
            the selection and reconstruction efficiency,
            the PID efficiency and the trigger efficiency,
            with each component determined separately.
            The geometrical acceptance, and the selection,
            reconstruction and trigger efficiencies are evaluated
            with the $p\mathrm{Pb}$ simulation samples.
            The simulation sample is weighted in order to
            match the occupancy of the tracking system observed in the data.
            The track reconstruction efficiency is calibrated
            with minimum-bias $\decay{\jpsi}{\mup\mun}$ and $\decay{\KS}{\pip\pim}$ samples,
            using the tag-and-probe approach employed in Ref.~\cite{LHCb-DP-2013-002}.
            The trigger efficiency obtained from the simulation is validated by measuring it from control data samples recorded with minimum trigger requirements.
            The PID efficiency is estimated with a tag-and-probe method~\cite{LHCb-PUB-2016-021,LHCb-DP-2018-001},
            using the \Dstar-tagged decay chain $\decay{\Dstarp}{\Dz \pip}$ with \dkpicf decays.


            Several sources of systematic uncertainty are considered
            and described in detail in the Supplemental Material~\cite{ref:SuppMat},
            where results and numerical values for the double-differential cross-section are also given.
            The total prompt \Dz production cross-section,
            obtained by integrating the double-differential measurements,
            is \mbox{$297.6 \pm 0.6 \pm14.0 \mbarn$} in the kinematic range of \mbox{$0<\pt<30\gevc$}
            and \mbox{$1.5<y^*<4.0$} for the forward rapidity region,
            and \mbox{$315.2 \pm 0.2 \pm17.8 \mbarn$}
            in the kinematic range of \mbox{$0<\pt<30\gevc$}
            and \mbox{$-5.0<y^*<-2.5$} in the backward rapidity region.
            The first uncertainties are statistical
            and the second systematic.


            The nuclear modification factor $R_{p\mathrm{Pb}}$ is defined as
            \begin{equation}
                R_{p\mathrm{Pb}}(\pt,y^*)\equiv \frac 1 A
                \frac{\deriv^2 \sigma_{p\text{Pb}}(\pt,y^*)/(\deriv\pt\deriv y^*)}
                {\deriv^2 \sigma_{pp}(\pt,y^*)/(\deriv\pt\deriv y^*)}~,
            \end{equation}
            where $A=208$ is the mass number of the lead nucleus and 
            $\sigma_{pp}$ is the prompt \Dz production cross-section
            in $pp$ collisions at $\sqs=8.16\tev$.
            An interpolation between
            \lhcb measurements at $\sqs = 5.02\tev$ and $\sqs=13\tev$~\cite{LHCb-PAPER-2016-042,LHCb-PAPER-2015-041}
            is performed to obtain
            $\deriv^2 \sigma_{pp}(\pt,y^*)/(\deriv\pt\deriv y^*)$,
            using a power-law function $\sigma(\sqs) = p_0 \left(\sqs\right)^{p_1}$.
            A linear function 
            is also considered. 
            The interpolation uncertainty comprises the difference between the two interpolation models, and the propagated total uncertainties from the $pp$ measurements, and typically amounts to 3\% (5\%) at forward (backward) rapidity.
            The interpolation is performed within the common measured kinematic range of $\pt < 10 \gevc$ and $2.0 < y < 4.5$
            for 5.02 and 13\tev $pp$ results,
            hence $R_{p\mathrm{Pb}}$ is measured in that range.

\begin{figure}[tb]
                \begin{center}
                    \includegraphics[width=0.98\linewidth]{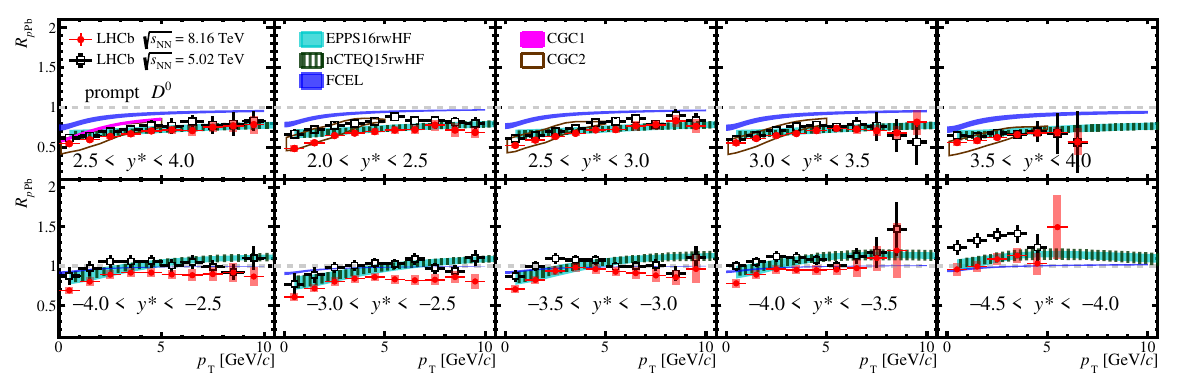}
                    \vspace*{-0.5cm}
                \end{center}
                \caption{\small
                Nuclear modification factor as a function of $\pt$
                in different $y^*$ intervals for prompt $\Dz$ mesons in the (top) forward and
                (bottom) backward regions.
                The error bars show the statistical uncertainties 
                and the boxes show the systematic uncertainties.
                The \lhcb results at $\sqsnn = 5.02\tev$~\cite{LHCb-PAPER-2017-015}
                and theoretical calculations at $\sqsnn=8.16\tev$ from
                Refs.~\cite{Eskola:2016oht,Kovarik:2015cma,Ducloue:2016ywt,Ma:2018bax,Arleo:2021bpv}
                are also shown.
                For \lhcb results at $\sqsnn=5.02\tev$,
                the error bars show the quadric sum of statistical and systematic uncertainties.
                }
                \label{fig:rpa_pt}
            \end{figure}

            The nuclear modification factor of the \Dz meson as a function of \pt
            is displayed in Fig.~\ref{fig:rpa_pt},
            where 8 panels report the results in different $y^*$ subintervals of $\Delta y^*=0.5$ and the two left panels are in the common range between the forward and backward rapidity coverage,  \mbox{$2.5<|y^*|<4$}.
            Figures showing $R_{p\mathrm{Pb}}$ and the forward-backward production ratio $R_{\mathrm{FB}}$ as functions of $y^*$ and \pt in $\Delta y^*=0.25$ intervals, as well as the numerical values are given in the Supplemental Material~\cite{ref:SuppMat}.
            A significant suppression of the cross-section in $p$Pb collisions,
            with respect to that in $pp$ collisions scaled by the lead mass number,
            is observed at forward rapidity as well as at backward rapidity up to $y^* \sim -3.5$.


            The $R_{p\mathrm{Pb}}$ results are compared with several theoretical calculations.
            The HELAC-Onia approach~\cite{Shao:2012iz,Shao:2015vga}
            is based on a data-driven modeling
            of the scattering at partonic level folded with
            the free proton PDFs~\cite{Lansberg:2016deg}.
            The calculations are first tuned by fitting the
            cross-sections measured in $pp$ collisions at the LHC.
            Then, the modified PDFs of nucleons in the Pb nucleus are introduced in the model
            to calculate the cross-sections in $p$Pb collisions and to estimate the effect of nPDFs,
            neglecting other cold and hot nuclear matter effects. 
            Reweighted EPPS16~\cite{Eskola:2016oht} and nCTEQ15~\cite{Kovarik:2015cma} nPDF sets, where LHC heavy flavor data~\cite{LHCb-PAPER-2017-015,ALICE:2014xjz,ALICE:2016cpm,ALICE:2016yta} are incorporated by performing a Bayesian-reweighting analysis~\cite{PhysRevLett.121.052004}, are used in the calculations, resulting in considerably reduced uncertainties than calculations using the default nPDFs. 
            The uncertainties are dominated by nPDF parameterizations
            and correspond to a 68\% confidence interval.
            At forward rapidity, the calculations are in general agreement with the data,
            except for $\pt<1\gevc$ where the predictions are about $2$ standard deviations larger than the data.
            This discrepancy suggests stronger shadowing or additional energy loss at low $x$.
            At backward rapidity, 
            for $\pt>6\gevc$ and $-3.5<y^*<-2.5$ the data are lower than the calculations by $2.0-3.8$ standard deviations,
            indicating a weaker antishadowing effect or possible final-state effects.


            The nuclear modification factor 
            is also compared with two calculations based on the CGC effective field theory, CGC1 and CGC2. 
            Since gluon saturation is expected to occur at small $x$ and $Q^2$, the calculations are applicable for $\pt <5\gevc$ at forward rapidity where saturation effects are relevant.
            For CGC1~\cite{Ducloue:2015gfa,Ducloue:2016ywt} the $D$-meson production is calculated with the color dipole formalism, 
            and the optical Glauber model is used to relate the initial condition of a nucleus to that of the proton.
            For CGC2~\cite{Ma:2018bax} the color dipole approach is combined with a heavy-quark fragmentation function to calculate the cross-sections.
            The CGC1 predictions have much smaller uncertainties than the CGC2 ones,
            because the CGC1 uncertainties include only variations of the $c$ quark mass and of the factorization scale,
            which largely cancel out in the $R_{p\mathrm{Pb}}$ ratio versus \pt.
            CGC1 is consistent with the upper bound of CGC2 and is slightly higher than the data.
            CGC2 shows a stronger suppression than HELAC-Onia calculations
            and gives a better description of the data, especially for $\pt<3\gevc$.

            A fourth calculation estimates \Dz suppression caused
            by medium-induced fully coherent energy loss (FCEL)~\cite{Arleo:2021bpv},
            a CNM effect where the interference between initial- and final-state gluon radiation results
            in an energy loss proportional to the incoming parton energy.
            The FCEL prediction shown in Fig.~\ref{fig:rpa_pt} does not consider the modification of nPDFs. 
            The effect is significant for low \pt,
            suggesting the suppression observed for $\pt<1\gevc$ may be caused by combined effects from nPDFs and FCEL.
            For $\pt>6\gevc$ the suppression due to FCEL is negligible,
            thus the discrepancy between the data and HELAC-Onia calculations with nPDFs at backward rapidity
            cannot be attributed to FCEL effects.

            The results are also compared with the \lhcb ~\Dz measurement at $\sqsnn = 5.02\tev$~\cite{LHCb-PAPER-2017-015}. At forward rapidity the $R_{p\mathrm{Pb}}$ values at the two energies are compatible, while at backward rapidity the 8.16\tev data are significantly lower. 
            The difference could be related to the different Bjorken-$x$ coverage at the two collision energies, while effects related to the Pb-going hemisphere other than nPDFs and FCEL, such as final-state energy loss in a high-particle-density environment, may also show a \sqsnn dependence as more charged hadrons are produced in 8.16\tev collisions.
            On the other hand, 
            the model calculations offer limited insights into collision energy dependence. 
            HELAC-Onia predictions based on nPDFs are compatible between the two \sqsnn values due to the large uncertainty of the nPDFs used in the 5.02\tev calculation. 
            The CGC models show similar values at 5.02 and 8.16\tev at forward rapidity while they are not applicable at backward. 
            Effects due to FCEL are generally small at backward rapidity.

            It is essential to study the impact of Bjorken-$x$ coverage in order to interpret the energy dependence observed in the data. However, $x$ and the momemtum transfer $Q^2$~\cite{Armesto:2006ph} are partonic quantities that cannot be directly measured in hadronic collisions. Instead, experimental proxies $x_\mathrm{exp}$ and $Q^2_\mathrm{exp}$, defined as
            %
            \begin{equation}
                x_\mathrm{exp} \equiv 2\frac{\sqrt{\pt^2(\Dz)+M^2(\Dz)}}{\sqsnn} e^{-y^*}~\text{and}~
                Q^2_\mathrm{exp} \equiv \pt^2(\Dz)+M^2(\Dz),
                \label{eqn:xandq2}
            \end{equation}
            are introduced to approximate the variation of $R_{p\mathrm{Pb}}$ with $x$ and $Q^2$, where $M(\Dz)$ and $\pt(\Dz)$ denote the mass and \pt of \Dz mesons, respectively.

\begin{figure}[tb]
                \begin{center}
                    \includegraphics[width=0.98\linewidth]{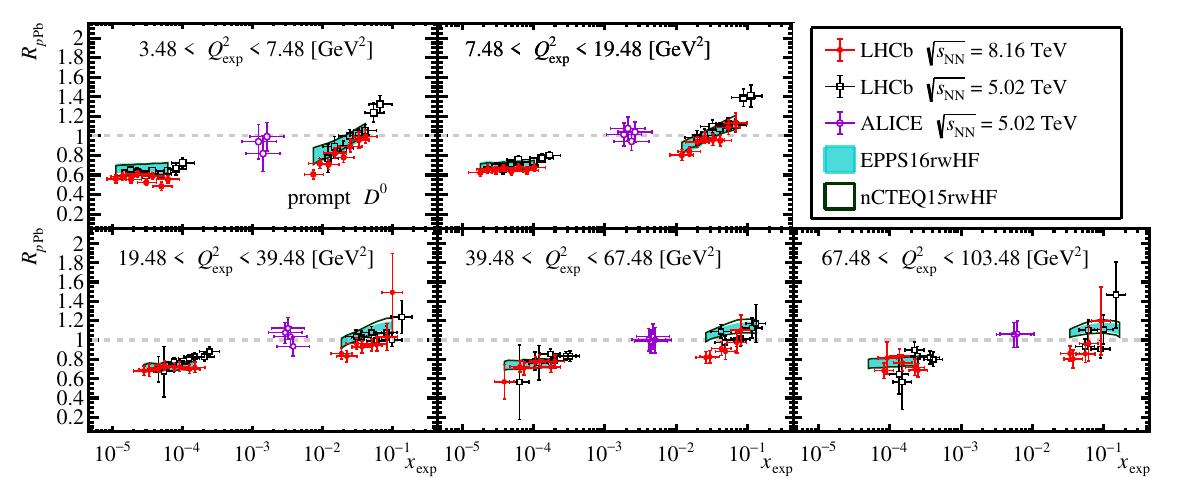}
                    \vspace*{-0.5cm}
                \end{center}
                \caption{\small
                Nuclear modification factor as a function of $x_\mathrm{exp}$
                in different $Q^2_\mathrm{exp}$ intervals for prompt $\Dz$ mesons
                for \lhcb results at $\sqsnn = 8.16\tev$ and $\sqsnn=5.02$~\cite{LHCb-PAPER-2017-015}
                and the \alice result at $\sqsnn=5.02\tev$~\cite{ALICE:2019fhe}.
                Theoretical calculations at $\sqsnn=8.16\tev$ from Refs.~\cite{Eskola:2016oht,Kovarik:2015cma}
                are also shown.
                The horizontal error bars account for the maximum and minimum
                $x_\mathrm{exp}$ values for a given $(\pt, y^*)$ interval
                and the vertical error bars show the quadric sum of statistical and systematic uncertainties.
                }
                \label{fig:rpa_x}
            \end{figure}

            Figure~\ref{fig:rpa_x} shows $R_{p\mathrm{Pb}}$ as a function of $x_\mathrm{exp}$
            in five $Q^2_\mathrm{exp}$ intervals, for \Dz mesons measured in this work at 8.16\tev, and at 5.02\tev from Ref.~\cite{LHCb-PAPER-2017-015}.
            The $x_\mathrm{exp}$ coverage of the 8.16\tev data extends lower than that of the 5.02\tev measurements
            due to the higher \sqsnn value, 
            reaching down to $x_\mathrm{exp} \sim 10^{-5}$ in the interval $3.48 < Q^2_\mathrm{exp} < 7.48\gev^2$, which corresponds to $\pt < 2\gevc$.
            The 8.16\tev data are also more precise.
            Data from the two energies are in good agreement with each other at common $x_\mathrm{exp}$ values.
            The measurements form a consistent trend
            from the small $x_\mathrm{exp}$ region corresponding to forward rapidity to the large $x_\mathrm{exp}$ region corresponding to backward rapidity,
            for all $Q^2_\mathrm{exp}$ intervals.
            The \Dz $R_{p\mathrm{Pb}}$ ratio at 5.02\tev at midrapidity~\cite{ALICE:2019fhe} measured by the \alice collaboration is also added to Fig.~\ref{fig:rpa_x},
            and is compatible with the trend within uncertainties.
            The trend suggests that the \sqsnn dependence observed at backward rapidity in Fig.~\ref{fig:rpa_pt} arises from different $x$ coverage in a kinematic region where $R_{p\mathrm{Pb}}$ depends strongly on $x$.

            The HELAC-Onia predictions are also transformed according to Eq.~\ref{eqn:xandq2} and shown in Fig.~\ref{fig:rpa_x}.
            In the small $x_\mathrm{exp}$ region, 
            the calculations are in general agreement with the data, 
            except for the interval $3.48 <Q^2_\mathrm{exp} < 7.48 \gev^2$ ($\pt < 2\gevc$) and $10^{-5} < x_\mathrm{exp} < 10^{-4}$, 
            where the nPDF expectations are slightly larger than the data and show greater uncertainty. 
            The data hint at a stronger shadowing effect, or other possible effects such as FCEL, that suppresses low-\pt ~\Dz production at forward rapidity.
            Moreover, estimations from Ref.~\cite{PhysRevLett.100.022303} suggest gluon saturation may occur in this region.
            At backward rapidity, 
            the $R_{p\mathrm{Pb}}$ values from the model are larger than those in the data for $Q^2_\mathrm{exp} > 19.48 \gev^2$ ($\pt > 4\gevc$) and $10^{-2} < x_\mathrm{exp} < 10^{-1}$, indicating smaller antishadowing effects in the data if nuclear effects other than nPDFs are negligible. Alternatively it suggests additional suppression mechanisms, such as final-state energy loss, may occur at backward rapidity.


            In summary, the prompt $\Dz$ production cross-section
            is measured at the \lhcb experiment in proton-lead collisions at
            $\sqsnn=8.16\tev$, at both forward and backward rapidities.
            The nuclear modification factors are measured with high accuracy and show strong cold nuclear matter effects.
            A stronger suppression than the predictions of nPDF calculations is observed for the lowest transverse momentum region of $\pt<1\gevc$
            at forward rapidity,
            hinting at a stronger shadowing than predicted at Bjorken-$x \sim 10^{-5}$, or additional effects at play.
            For the backward rapidity range of $-3.5<y^*<-2.5$,
            the $R_{p\mathrm{Pb}}$ values are lower than nPDF calculations
            at $\pt>6 \gevc$ with a significance of $2.0-3.8$ standard deviations,
            indicating 
            a weaker antishadowing effect than the model
            or additional final-state effects at backward rapidity.
            This Letter presents the most precise measurement of the prompt \Dz production
            in $p$Pb collisions to date, 
            providing unique constraints to improve nPDF parameterization down to $x \sim 10^{-5}$.
            \clearpage

            \input{supplementary}

            \clearpage

            \section*{Acknowledgements}
            \noindent We express our gratitude to our colleagues in the CERN
            accelerator departments for the excellent performance of the LHC. We
            thank the technical and administrative staff at the LHCb
            institutes.
            We acknowledge support from CERN and from the national agencies:
            CAPES, CNPq, FAPERJ and FINEP (Brazil); 
            MOST and NSFC (China); 
            CNRS/IN2P3 (France); 
            BMBF, DFG and MPG (Germany); 
            INFN (Italy); 
            NWO (Netherlands); 
            MNiSW and NCN (Poland); 
            MEN/IFA (Romania); 
            MICINN (Spain); 
            SNSF and SER (Switzerland); 
            NASU (Ukraine); 
            STFC (United Kingdom); 
            DOE NP and NSF (USA).
            We acknowledge the computing resources that are provided by CERN, IN2P3
            (France), KIT and DESY (Germany), INFN (Italy), SURF (Netherlands),
            PIC (Spain), GridPP (United Kingdom), 
            CSCS (Switzerland), IFIN-HH (Romania), CBPF (Brazil),
            Polish WLCG  (Poland) and NERSC (USA).
            We are indebted to the communities behind the multiple open-source
            software packages on which we depend.
            Individual groups or members have received support from
            ARC and ARDC (Australia);
            Minciencias (Colombia);
            AvH Foundation (Germany);
            EPLANET, Marie Sk\l{}odowska-Curie Actions and ERC (European Union);
            A*MIDEX, ANR, IPhU and Labex P2IO, and R\'{e}gion Auvergne-Rh\^{o}ne-Alpes (France);
            Key Research Program of Frontier Sciences of CAS, CAS PIFI, CAS CCEPP, 
            Fundamental Research Funds for the Central Universities, 
            and Sci. \& Tech. Program of Guangzhou (China);
            GVA, XuntaGal, GENCAT and Prog. Atracci\'on Talento, CM (Spain);
            SRC (Sweden);
            the Leverhulme Trust, the Royal Society
            and UKRI (United Kingdom).

            \addcontentsline{toc}{section}{References}
            \ifx\mcitethebibliography\mciteundefinedmacro
            \PackageError{LHCb.bst}{mciteplus.sty has not been loaded}
            {This bibstyle requires the use of the mciteplus package.}\fi
            \providecommand{\href}[2]{#2}

            \clearpage
            \input{Authorship_LHCb-PAPER-2022-007}

        \end{document}

%% file: supplementary.tex
\def\logipchisq{\ensuremath{\log_{10}\left(\chi^2_{\text{IP}}\right)}}
\section*{Supplemental Material}
\label{sec:Supplementary}
\subsection*{Fit to $M(\kaon\pion)$ and \logipchisq\ distributions}
\begin{figure}[htbp]
    \begin{center}
    \includegraphics[width=0.9\linewidth]{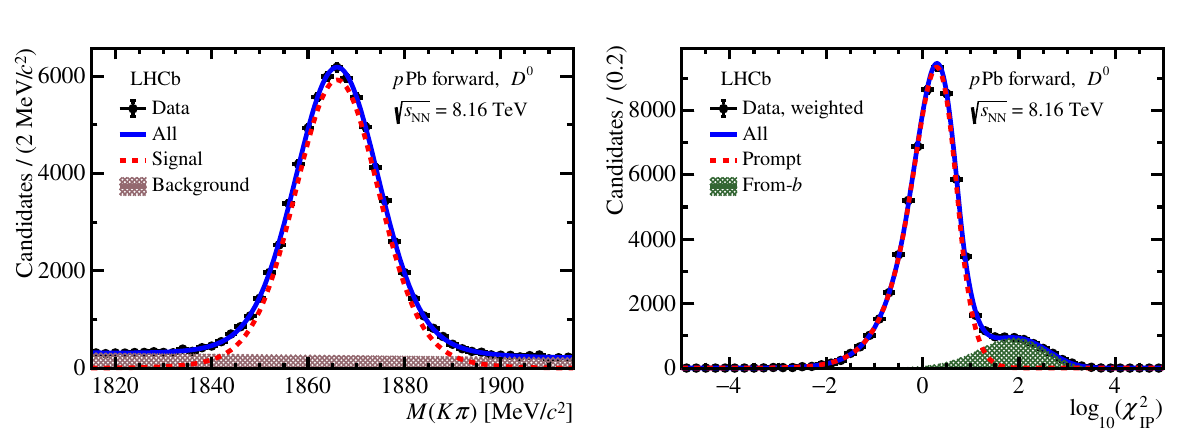}
        \vspace*{-0.5cm}
    \end{center}
    \caption{\small
    Distributions and fit results of (left) $M(\kaon\pion)$ and (right) \logipchisq\
    for inclusive \Dz mesons in the forward data sample in the kinematic range
    of \mbox{$2.5<\pt<3.0\gevc$} and \mbox{$3.25<y^*<3.50$}.
    For the \logipchisq\ fit, the data are weighted using the \sPlot method to subtract the background component.
    }
    \label{fig:fit1}
\end{figure}

\begin{figure}[htbp]
    \begin{center}
        \includegraphics[width=0.9\linewidth]{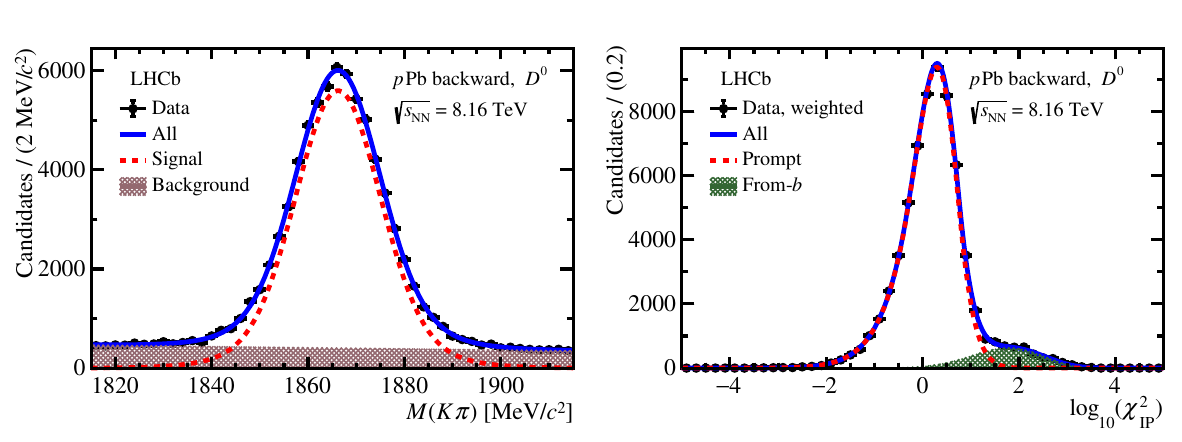}
        \vspace*{-0.5cm}
    \end{center}
    \caption{\small
    Distributions and fit results of (left) $M(\kaon\pion)$ and (right) \logipchisq\
    for inclusive \Dz mesons in the backward data sample
    in the kinematic range of \mbox{$2.5<\pt<3.0\gevc$} and \mbox{$-4.50<y^*<-4.25$}.
    For the \logipchisq\ fit, the data are weighted using the \sPlot method to subtract the background component.
    }
    \label{fig:fit2}
\end{figure}

The fit results for the $M(\kaon\pion)$ and \logipchisq\ distributions
in the kinematic ranges of $2.5<\pt<3.0\gevc$ and
$3.25<y^*<3.50$ $(-4.50<y^*<-4.25)$
are shown in Fig.~\ref{fig:fit1} (Fig.~\ref{fig:fit2}).
The $M(\kaon \pion)$ distribution of the signal is described by a sum
of a Crystal Ball function~\cite{Skwarnicki:1986xj} and a Gaussian function sharing a common mean value. The background shape is described by a linear function.
The shapes of the \logipchisq\ distribution
corresponding to the prompt and \mbox{\it from-\bquark} components
are described independently by Bukin functions~\cite{bukin2007fitting}.

\subsection*{
Systematic uncertainties}
Several sources of systematic uncertainty are considered
and are evaluated separately for the forward and backward samples, unless stated otherwise.
The systematic uncertainty due to the invariant mass $M(\kaon\pion)$ modelling is studied using alternative models. The sum of two Crystal Ball functions~\cite{Skwarnicki:1986xj} is used to describe the signals, and an exponential function used to describe the background.
For the fit to the \logipchisq\ distribution, the parameters describing the tail of the Bukin functions~\cite{bukin2007fitting}
and the parameter describing the asymmetry of the {\it from-\bquark} component,
which are fixed to values derived from the simulation in the default result,
are varied by one standard deviation. 
The effect due to the {\it from-\bquark} component modelling is studied by substituting the nominal Bukin function with a Gaussian function.
Finally, the background subtraction with the \sPlot technique~\cite{pivk2005plots} is checked by performing the fit with the method described in Ref.~\cite{LHCb-PAPER-2017-015}, where the background is constructed with candidates in the side band.
In these tests, the largest difference from the default value is considered as the systematic uncertainty.
The uncertainties of the tracking and PID calibration
are dominated by those arising from the statistically limited calibration sample size.
The uncertainty arising from matching the simulated detector occupancy distribution
to the data is estimated by weighting with different variables.
For the trigger efficiency, the difference between the efficiencies derived from the simulation
and from collision data~\cite{LHCb-PUB-2014-039} is considered as the uncertainty.
The uncertainties arising from the luminosity,
the $\mathcal{B}(\decay{\Dz}{\Kmp\pipm})$ branching fractions,
the statistically limited simulation sample size and the uncertainty
from the $pp$ interpolation are also included.
The ranges of systematic uncertainties for the forward and backward rapidity regions
are listed in Table~\ref{tab:sys}.

\begin{table}[htbp]
    \centering
    \caption{Systematic uncertainties considered in this measurement, in \%.
    The range indicates the minimum and the maximum value among the two-dimensional \pt and $y^*$ intervals.
    The systematic uncertainties due to simulation sample size,
    mass fit and \logipchisq\ fit are uncorrelated across the intervals.
    The other sources of systematic uncertainties are fully correlated between different intervals.
    }
    \begin{tabular}{l|cc}
        \hline
        Uncertainty source & Forward [\%]  & Backward [\%]\\
        \hline
        Tracking calibration & 3.0~--~4.7 & 3.1~--~10.7  \\
        PID & 0.2~--~6.9 & 0.2~--~26.5 \\
        Trigger efficiency & 0.0~--~16.5 & 0.0~--~4.7 \\
        Multiplicity correction & 0~--~9 & 0~--~16 \\
        Luminosity & 2.6 & 2.5 \\
        Branching fraction & 0.8 & 0.8 \\
        Mass fit & 0.0~--~19.3 & 0.1~--~6.1 \\
        \logipchisq\ fit & 0.3~--~19.5 & 0.4~--~7.0 \\
        Simulation sample size & 1~--~40 & 1~--~26 \\
        $pp$ interpolation & 3.4~--~17.5 & 3.4~--~28.8 \\
        \hline
    \end{tabular}\label{tab:sys}
\end{table}
\subsection*{Double differential cross-section}
\begin{figure}[tb]
    \begin{center}
        \includegraphics[width=0.8\linewidth]{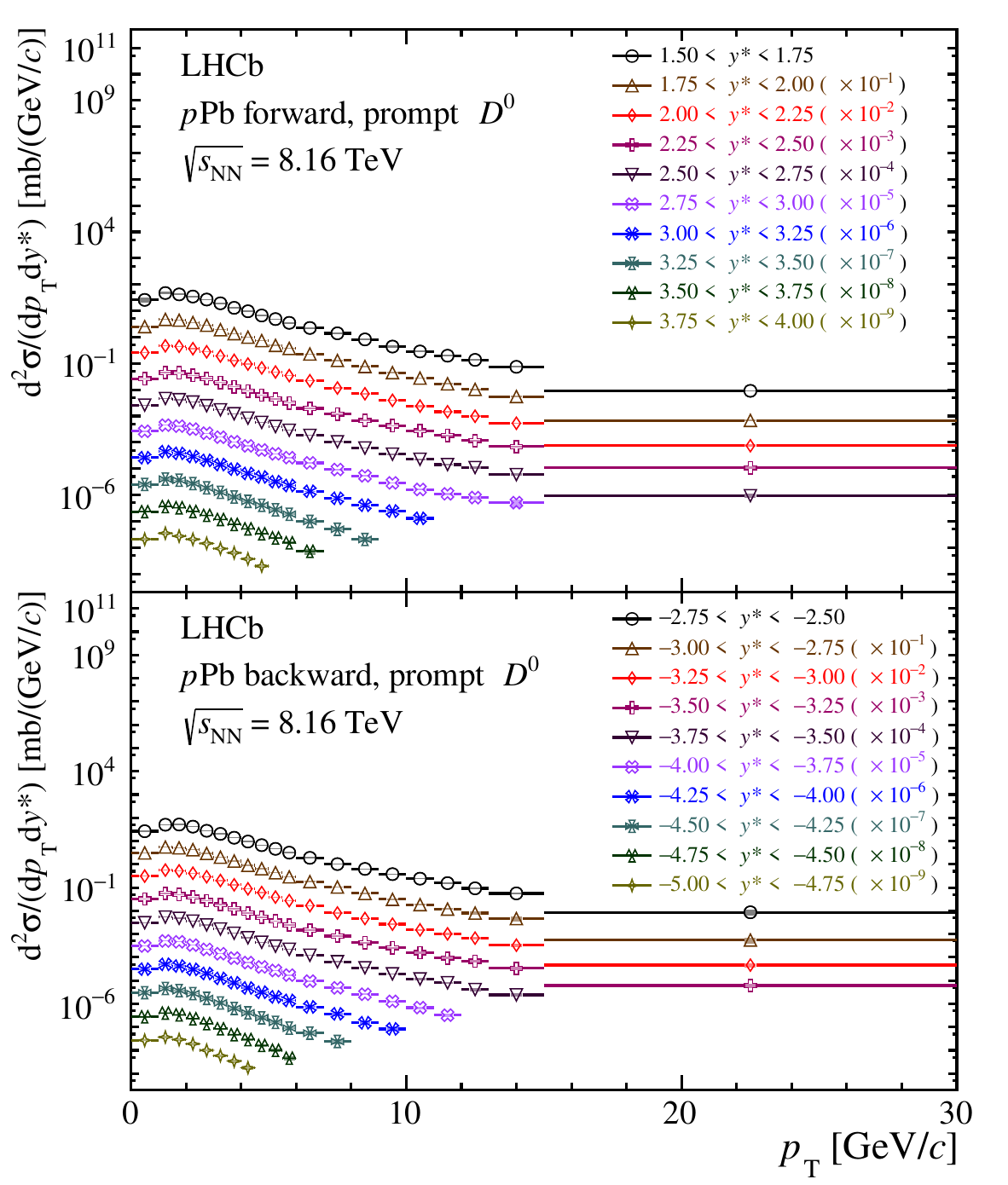}
        \vspace*{-0.5cm}
    \end{center}
    \caption{\small
    Double-differential cross-sections of prompt \Dz mesons in $p$Pb collisions
    in the (top) forward and (bottom) backward rapidity regions.
    To display the differential \mbox{cross-section} values in different rapidity intervals,
    multiplicative factors of $10^{-n}$ are used with $n$ increasing with rapidity value.
    The uncertainties are smaller than the symbol size.
    }
    \label{fig:cross_section}
\end{figure}

The double differential cross-sections for prompt \Dz mesons
in both forward and backward rapidity regions are shown in Fig.~\ref{fig:cross_section}.

\subsection*{Nuclear modification factor in different $y^*$ with \mbox{$\Delta y^*=0.25$} intervals }
\begin{figure}[htb]
    \begin{center}
        \includegraphics[width=0.98\linewidth]{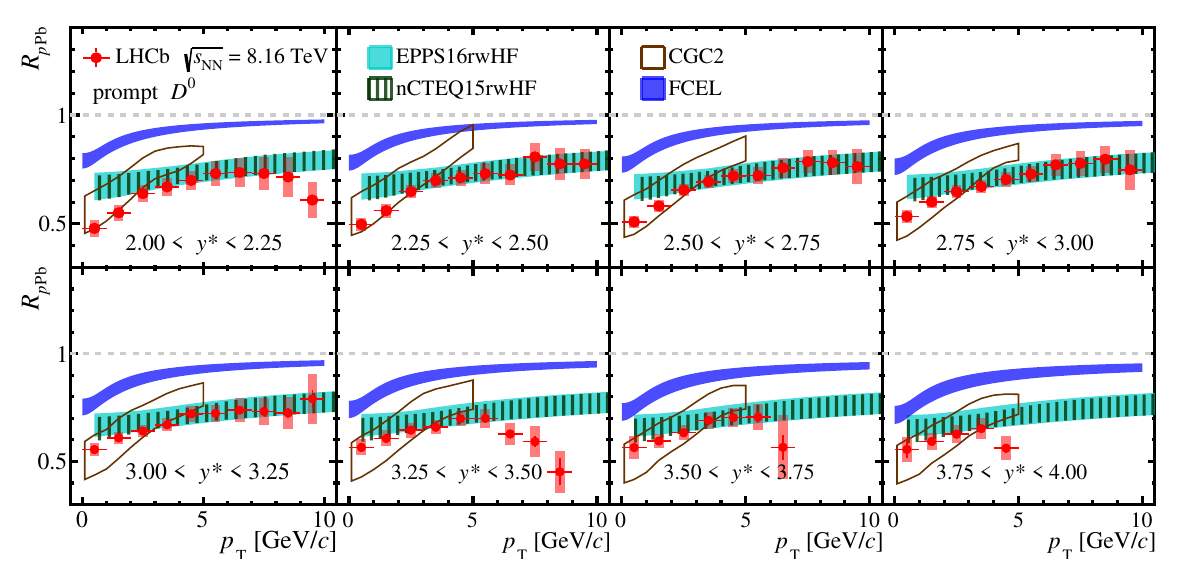}
        \vspace*{-0.5cm}
    \end{center}
    \caption{\small
    Nuclear modification factor as a function of $\pt$
    in different $y^*$ intervals for prompt $\Dz$ mesons in the forward regions
    for $2.0<y^*<4.0$.
    The error bars show the statistical uncertainties 
    and the boxes show the systematic uncertainties.
    The theoretical calculations from
    Refs.~\cite{Eskola:2016oht,Kovarik:2015cma,Ma:2018bax,Arleo:2021bpv}
    are also shown.
    }
    \label{fig:rpa_tight1}
\end{figure}

\begin{figure}[htb]
    \begin{center}
        \includegraphics[width=0.98\linewidth]{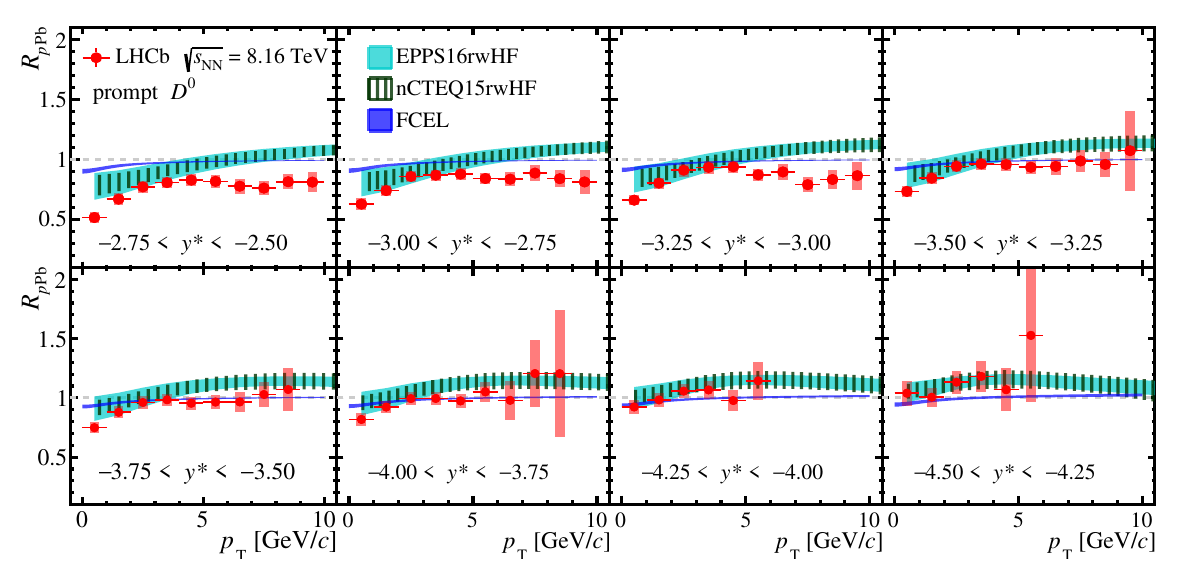}
        \vspace*{-0.5cm}
    \end{center}
    \caption{\small
    Nuclear modification factor as a function of $\pt$
    in different $y^*$ intervals for prompt $\Dz$ mesons in the backward regions
    for $-4.5<y^*<-2.5$.
    The error bars show the statistical uncertainties 
    and the boxes show the systematic uncertainties.
    The theoretical calculations from
    Refs.~\cite{Eskola:2016oht,Kovarik:2015cma,Arleo:2021bpv}
    are also shown.
    }
    \label{fig:rpa_tight2}
\end{figure}

The nuclear modification factor for the forward rapidity regions of
$2.0<y^*<4.0$ are shown in Fig.~\ref{fig:rpa_tight1}
while the backward regions of $-4.5<y^*<-2.5$ are shown in Fig.~\ref{fig:rpa_tight2}.
The HELAC-Onia calculation~\cite{Shao:2012iz,Shao:2015vga}
incorporating reweighted EPPS16~\cite{Eskola:2016oht} and nCTEQ15~\cite{Kovarik:2015cma} nPDF sets,
as well as the fully coherent energy loss model~\cite{Arleo:2021bpv} without considering the modification of nPDFs,
are included.
Two color glass condensate predictions~\cite{Ducloue:2016ywt,Ma:2018bax}
are also shown at forward rapidity.


\subsection*{Nuclear modification factor as a function of $y^*$ intervals in full-\pt and high-\pt regions}
\begin{figure}[tb]
    \begin{center}
        \includegraphics[width=0.98\linewidth]{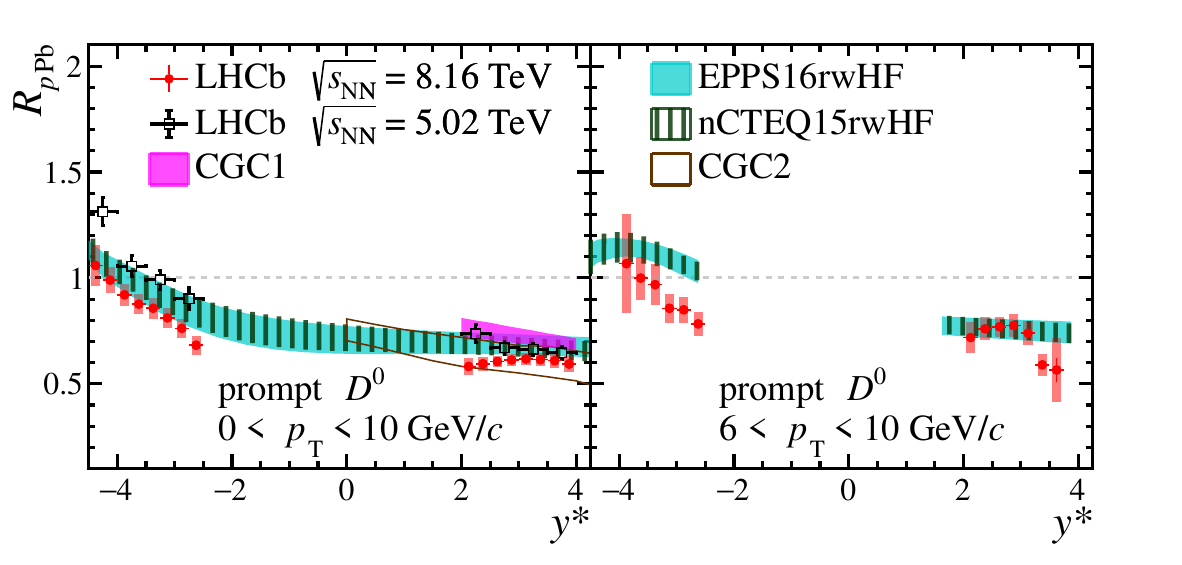}
        \vspace*{-0.5cm}
    \end{center}
    \caption{\small
    Nuclear modification factor for prompt $\Dz$ mesons as a function of $y^*$
    in (left) the full-\pt range and (right) the high-\pt range.
    The error bars show the statistical uncertainties 
    and the boxes show the systematic uncertainties.
    The \lhcb results at $\sqsnn=5.02\tev$~\cite{LHCb-PAPER-2017-015} and theoretical calculations at $\sqsnn=8.16\tev$ from
    Refs.~\cite{Eskola:2016oht,Kovarik:2015cma,Ducloue:2016ywt,Ma:2018bax}
    are also shown.
    For the \lhcb results at $\sqsnn=5.02\tev$,
    the error bars show the quadratic sum of statistical and systematic uncertainties.
    On the left,
    the \pt range is $0 < \pt< 15\gevc$ for the calculations with nPDFs of EPPS16 and nCTEQ15.
    }
    \label{fig:rpa_y}
\end{figure}

The nuclear modification factor as a function of $y^*$ in full-\pt $(0<\pt<10\gevc)$ and high-\pt regions $(6<\pt<10\gevc)$
are shown in Fig.~\ref{fig:rpa_y}.

\subsection*{Forward-backward production ratio $R_\mathrm{FB}$}
\begin{figure}[tb]
    \begin{center}
        \includegraphics[width=0.98\linewidth]{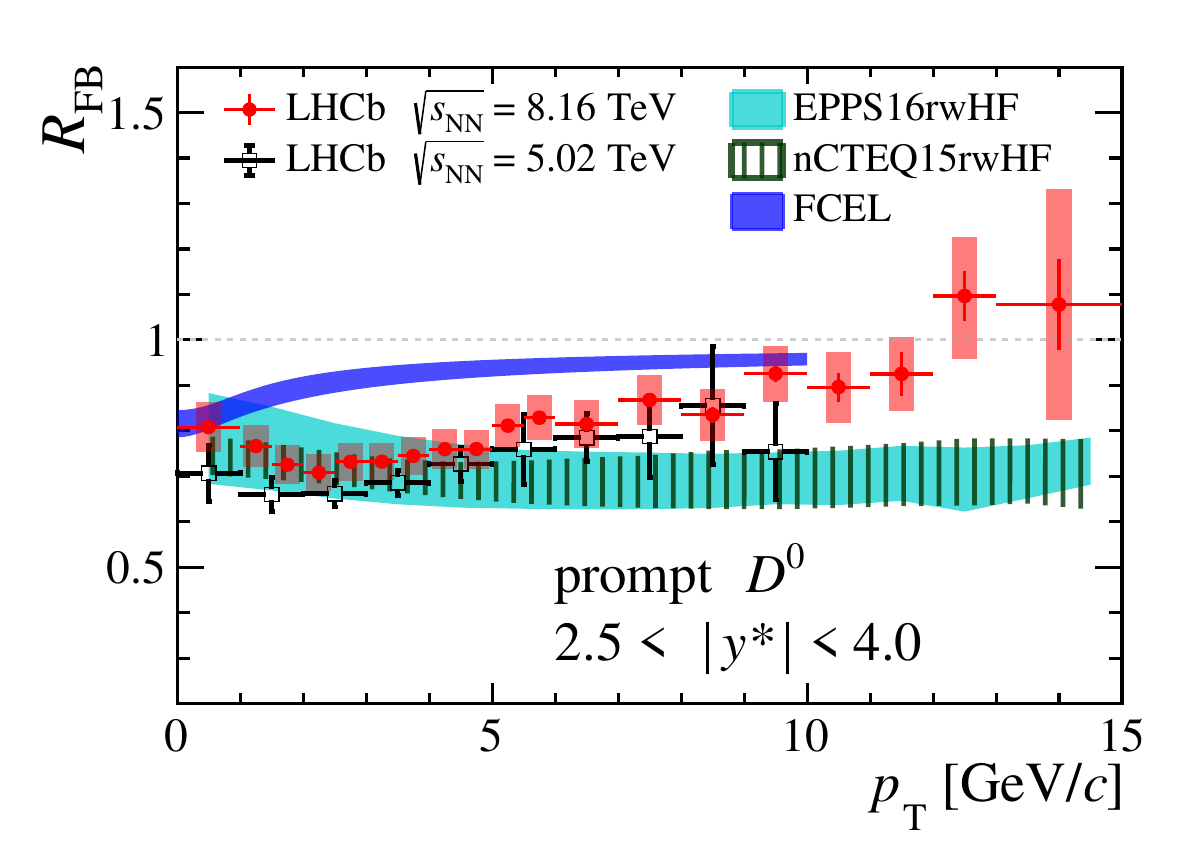}
        \vspace*{-0.5cm}
    \end{center}
    \caption{\small
    Forward-backward production ratio
    for prompt \Dz mesons as a function of \pt,
    integrated over the common rapidity range $2.5 < |y^*| < 4.0$.
    The error bars show the \mbox{statistical} uncertainties
    and the boxes show the systematic uncertainties.
    The \lhcb results at \mbox{$\sqsnn = 5.02\tev$}~\cite{LHCb-PAPER-2017-015}
    and theoretical calculations at $\sqsnn=8.16\tev$ from
    Refs.~\cite{Eskola:2016oht,Kovarik:2015cma,Arleo:2021bpv}
    are also shown.
    For the \lhcb results at $\sqsnn=5.02\tev$,
    the error bars show the quadratic sum of statistical and systematic uncertainties.
    }
    \label{fig:rfb_ptonly}
\end{figure}

\begin{figure}[htbp]
    \begin{center}
        \includegraphics[width=0.98\linewidth]{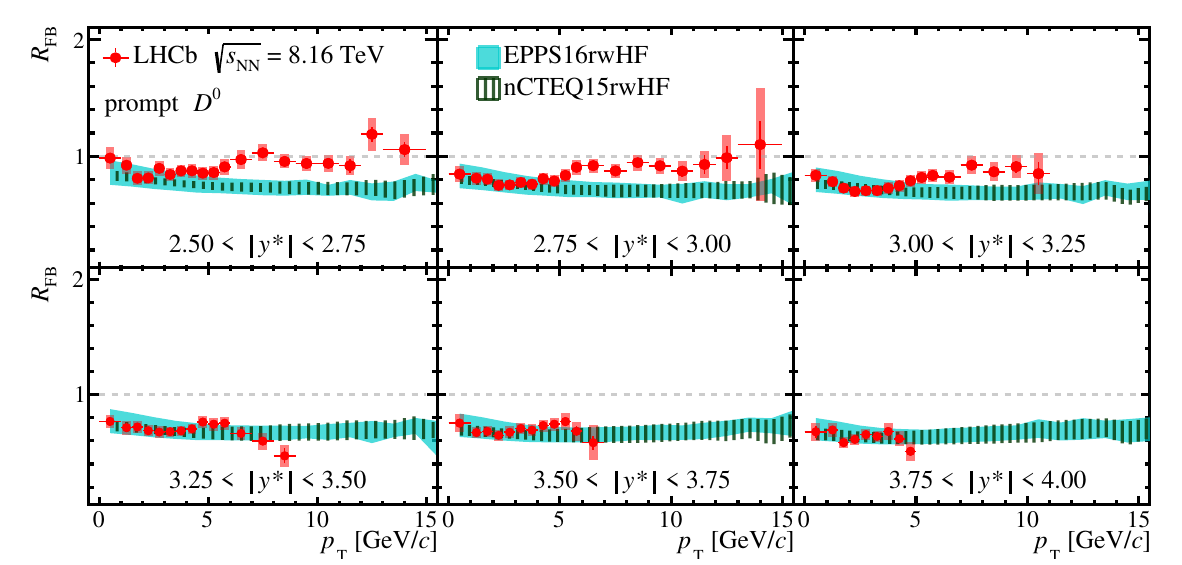}
        \vspace*{-0.5cm}
    \end{center}
    \caption{\small
    Forward and backward production ratio $R_{\mathrm{FB}}$ 
    for prompt \Dz mesons as a function of \pt
    in different $|y^*|$ intervals.
    The error bars show the statistical uncertainties 
    and the boxes show the systematic uncertainties.
    The theoretical calculations from
    Refs.~\cite{Eskola:2016oht,Kovarik:2015cma}
    are also shown.
    \label{fig:rfb_pt}
    }
\end{figure}

The forward-backward production ratio is defined as
\begin{equation}
    R_\mathrm{FB}(\pt, y^*) \equiv \frac
    {\deriv^2 \sigma(\pt,|y^*|;y^*>0)/(\deriv \pt \deriv y^*)}
    {\deriv^2 \sigma(\pt,|y^*|;y^*<0)/(\deriv \pt \deriv y^*)}~,
\end{equation}
which is calculated in the common $|y^*|$ interval 
of the forward-backward acceptance, $2.5<|y^*|<4$.
The measurements of $R_\mathrm{FB}$ are
shown as a function of \pt in Fig.~\ref{fig:rfb_ptonly},
along with the \lhcb\ $\sqsnn = 5.02\tev$ results~\cite{LHCb-PAPER-2017-015}
and the nPDF calculations.
The numerical values for $R_\mathrm{FB}$ are given
in Table~\ref{tab:RFB_pt}.
Measurements of $R_\mathrm{FB}$~\vs\pt $|y^*|$ intervals with $\Delta|y^*|=0.25$
are shown in Fig.~\ref{fig:rfb_pt}.
Good agreement with nPDF calculations is found at low \pt. However, the data show a clear rising trend with increasing \pt, reaching unity at the highest \pt values, in contrast to the nPDF calculations, which predict $R_\mathrm{FB} \sim 0.7$ almost independent of \pt. This difference originates from the suppression of high-\pt~\Dz mesons at backward rapidity.


\subsection*{Tables of numerical values for the double-differential cross-section,
nuclear modification factor and forward-backward ratio}
    The numerical values for the double-differential cross-section $\deriv^2 \sigma/(\deriv \pt \deriv y^*)$
    are listed in Tables~\ref{tab:cross_2D1} and \ref{tab:cross_2D2}
    for the forward and backward rapidity regions.
    The numerical values for the nuclear modification factor $R_{p\mathrm{Pb}}$
    are listed in Tables~\ref{tab:RpA_loose1} and \ref{tab:RpA_loose2}
    for the forward and backward rapidity regions.
    The numerical values for the forward-backward ratio $R_{\mathrm{FB}}$
    as a function of $\pt$ integrated in the rapidity region of $2.5<|y^*|<4.0$
    are listed in Table~\ref{tab:RFB_pt}.

\begin{sidewaystable}[htbp] 
	\caption{Double-differential cross-sections for prompt \Dz mesons in intervals of \pt and $y^*$
    in forward rapidity regions.
    The first uncertainty is statistical,
    the second is the component of the systematic uncertainty
    that is uncorrelated across intervals and the third is the correlated component.} 
	\centering 
	\scalebox{0.7}{ 
	\begin{tabular}{cccccc} 
		\hline 
		\multicolumn{6}{c}{$\deriv^2 \sigma/(\deriv \pt \deriv y)~[\mathrm{mb}/(\gevc)]$} \\ 
		$\pt~[\gevc]\backslash y^*$	&(1.50, 1.75)	&(1.75, 2.00)	&(2.00, 2.25)	&(2.25, 2.50)	&(2.50, 2.75) \\ 
		\hline 
		(0.0,1.0)	&$26.049\pm1.734\pm1.961\pm4.494$	&$25.059\pm0.128\pm0.594\pm1.304$	&$25.716\pm0.090\pm0.327\pm1.205$	&$26.194\pm0.079\pm0.455\pm1.208$	&$26.046\pm0.075\pm0.607\pm1.162$ \\ 
		(1.0,1.5)	&$47.452\pm0.229\pm2.317\pm4.304$	&$47.375\pm0.222\pm0.870\pm2.639$	&$46.707\pm0.160\pm0.630\pm2.243$	&$45.091\pm0.137\pm1.369\pm2.000$	&$46.465\pm0.138\pm0.780\pm1.999$ \\ 
		(1.5,2.0)	&$42.066\pm0.153\pm1.760\pm2.658$	&$44.164\pm0.188\pm0.962\pm2.054$	&$44.574\pm0.103\pm0.564\pm1.952$	&$43.939\pm0.129\pm0.876\pm2.220$	&$41.657\pm0.120\pm0.852\pm1.793$ \\ 
		(2.0,2.5)	&$34.993\pm0.350\pm1.535\pm1.662$	&$35.978\pm0.140\pm0.653\pm1.709$	&$36.344\pm0.112\pm0.546\pm1.558$	&$35.023\pm0.103\pm0.550\pm1.541$	&$33.970\pm0.095\pm0.659\pm1.413$ \\ 
		(2.5,3.0)	&$26.858\pm0.227\pm0.865\pm1.295$	&$28.324\pm0.109\pm0.440\pm1.250$	&$27.658\pm0.089\pm0.358\pm1.336$	&$26.622\pm0.080\pm0.554\pm1.151$	&$24.771\pm0.095\pm0.536\pm1.062$ \\ 
		(3.0,3.5)	&$19.321\pm0.165\pm0.615\pm0.963$	&$19.133\pm0.155\pm0.295\pm0.817$	&$19.185\pm0.062\pm0.381\pm0.828$	&$18.717\pm0.069\pm0.251\pm0.822$	&$17.124\pm0.076\pm0.345\pm0.737$ \\ 
		(3.5,4.0)	&$12.973\pm0.114\pm0.409\pm0.563$	&$13.543\pm0.057\pm0.282\pm0.578$	&$13.060\pm0.059\pm0.336\pm0.672$	&$12.918\pm0.057\pm0.354\pm0.542$	&$12.153\pm0.056\pm0.248\pm0.531$ \\ 
		(4.0,4.5)	&$9.619\pm0.098\pm0.379\pm0.449$	&$9.947\pm0.120\pm0.222\pm0.432$	&$9.450\pm0.055\pm0.189\pm0.405$	&$8.918\pm0.057\pm0.234\pm0.386$	&$8.350\pm0.038\pm0.205\pm0.360$ \\ 
		(4.5,5.0)	&$6.639\pm0.075\pm0.323\pm0.341$	&$7.358\pm0.039\pm0.205\pm0.323$	&$6.747\pm0.036\pm0.137\pm0.292$	&$6.269\pm0.042\pm0.170\pm0.267$	&$5.691\pm0.036\pm0.145\pm0.239$ \\ 
		(5.0,5.5)	&$4.851\pm0.063\pm0.194\pm0.212$	&$4.857\pm0.040\pm0.151\pm0.215$	&$4.723\pm0.032\pm0.114\pm0.200$	&$4.430\pm0.035\pm0.126\pm0.194$	&$3.983\pm0.028\pm0.105\pm0.177$ \\ 
		(5.5,6.0)	&$3.492\pm0.051\pm0.163\pm0.148$	&$3.726\pm0.027\pm0.088\pm0.168$	&$3.437\pm0.029\pm0.088\pm0.146$	&$3.142\pm0.025\pm0.121\pm0.132$	&$2.877\pm0.022\pm0.087\pm0.130$ \\ 
		(6.0,7.0)	&$2.230\pm0.023\pm0.104\pm0.099$	&$2.275\pm0.022\pm0.058\pm0.097$	&$2.196\pm0.019\pm0.061\pm0.097$	&$1.964\pm0.014\pm0.053\pm0.086$	&$1.842\pm0.013\pm0.049\pm0.081$ \\ 
		(7.0,8.0)	&$1.395\pm0.021\pm0.063\pm0.060$	&$1.297\pm0.013\pm0.037\pm0.056$	&$1.161\pm0.014\pm0.044\pm0.050$	&$1.188\pm0.009\pm0.049\pm0.053$	&$1.052\pm0.009\pm0.032\pm0.048$ \\ 
		(8.0,9.0)	&$0.818\pm0.013\pm0.022\pm0.036$	&$0.764\pm0.010\pm0.018\pm0.032$	&$0.702\pm0.008\pm0.014\pm0.031$	&$0.682\pm0.008\pm0.022\pm0.031$	&$0.605\pm0.007\pm0.015\pm0.028$ \\ 
		(9.0,10.0)	&$0.438\pm0.019\pm0.032\pm0.020$	&$0.446\pm0.007\pm0.009\pm0.019$	&$0.394\pm0.006\pm0.012\pm0.018$	&$0.428\pm0.006\pm0.010\pm0.020$	&$0.355\pm0.006\pm0.009\pm0.018$ \\ 
		(10.0,11.0)	&$0.284\pm0.016\pm0.015\pm0.013$	&$0.278\pm0.007\pm0.010\pm0.012$	&$0.239\pm0.005\pm0.007\pm0.028$	&$0.276\pm0.005\pm0.008\pm0.014$	&$0.227\pm0.005\pm0.008\pm0.012$ \\ 
		(11.0,12.0)	&$0.196\pm0.009\pm0.008\pm0.008$	&$0.168\pm0.005\pm0.007\pm0.008$	&$0.145\pm0.004\pm0.005\pm0.007$	&$0.184\pm0.005\pm0.009\pm0.011$	&$0.143\pm0.005\pm0.006\pm0.008$ \\ 
		(12.0,13.0)	&$0.136\pm0.005\pm0.006\pm0.006$	&$0.103\pm0.005\pm0.008\pm0.005$	&$0.098\pm0.003\pm0.004\pm0.005$	&$0.116\pm0.004\pm0.006\pm0.007$	&$0.110\pm0.004\pm0.006\pm0.006$ \\ 
		(13.0,15.0)	&$0.074\pm0.004\pm0.004\pm0.004$	&$0.055\pm0.002\pm0.003\pm0.002$	&$0.053\pm0.002\pm0.002\pm0.003$	&$0.069\pm0.002\pm0.003\pm0.005$	&$0.059\pm0.003\pm0.004\pm0.004$ \\ 
        (15.0,30.0)	&$(9.1\pm0.5\pm0.5\pm0.5)\times 10^{-3}$    &$(6.8\pm0.3\pm0.3\pm0.3)\times 10^{-3}$    &$(7.4\pm0.3\pm0.4\pm0.8)\times 10^{-3}$    &$(10.7\pm0.5\pm0.7\pm0.9) \times 10^{-3}$    &$(9.3\pm1.1\pm1.3\pm0.7)\times 10^{-3}$ \\
		\hline 
		$\pt [\gevc]\backslash y^*$	&(2.75, 3.00)	&(3.00, 3.25)	&(3.25, 3.50)	&(3.50, 3.75)	&(3.75, 4.00) \\ 
		\hline 
		(0.0,1.0)	&$26.300\pm0.149\pm0.840\pm1.169$	&$26.038\pm0.083\pm0.662\pm1.191$	&$24.968\pm0.147\pm1.061\pm1.189$	&$23.157\pm0.196\pm1.842\pm1.157$	&$20.817\pm1.548\pm1.918\pm1.093$ \\ 
		(1.0,1.5)	&$44.909\pm0.177\pm1.057\pm1.965$	&$44.107\pm0.152\pm0.649\pm1.959$	&$39.851\pm0.269\pm2.554\pm1.807$	&$36.184\pm0.336\pm1.706\pm1.688$	&$35.092\pm0.401\pm1.848\pm1.984$ \\ 
		(1.5,2.0)	&$41.222\pm0.124\pm1.261\pm1.769$	&$37.934\pm0.172\pm0.827\pm1.668$	&$36.200\pm0.203\pm1.124\pm1.708$	&$32.475\pm0.262\pm1.177\pm1.541$	&$27.126\pm0.421\pm1.168\pm1.492$ \\ 
		(2.0,2.5)	&$31.966\pm0.121\pm0.963\pm1.501$	&$28.923\pm0.127\pm0.644\pm1.291$	&$27.061\pm0.139\pm0.855\pm1.342$	&$23.492\pm0.139\pm0.796\pm1.048$	&$20.558\pm0.243\pm1.053\pm1.147$ \\ 
		(2.5,3.0)	&$22.247\pm0.087\pm0.451\pm0.977$	&$20.647\pm0.087\pm0.412\pm0.918$	&$18.474\pm0.097\pm0.814\pm0.812$	&$16.646\pm0.123\pm0.755\pm0.779$	&$14.379\pm0.208\pm0.637\pm0.811$ \\ 
		(3.0,3.5)	&$15.693\pm0.070\pm0.332\pm0.708$	&$14.210\pm0.066\pm0.308\pm0.631$	&$12.554\pm0.072\pm0.365\pm0.573$	&$11.648\pm0.101\pm0.451\pm0.563$	&$9.117\pm0.152\pm0.424\pm0.501$ \\ 
		(3.5,4.0)	&$10.468\pm0.052\pm0.246\pm0.459$	&$9.542\pm0.048\pm0.260\pm0.425$	&$8.336\pm0.055\pm0.285\pm0.370$	&$7.540\pm0.072\pm0.303\pm0.409$	&$6.285\pm0.185\pm0.525\pm0.406$ \\ 
		(4.0,4.5)	&$7.317\pm0.038\pm0.159\pm0.317$	&$6.576\pm0.037\pm0.174\pm0.304$	&$5.675\pm0.043\pm0.155\pm0.258$	&$5.108\pm0.068\pm0.204\pm0.253$	&$3.678\pm0.138\pm0.273\pm0.219$ \\ 
		(4.5,5.0)	&$5.108\pm0.030\pm0.116\pm0.234$	&$4.799\pm0.033\pm0.136\pm0.216$	&$4.029\pm0.040\pm0.121\pm0.188$	&$3.322\pm0.061\pm0.149\pm0.167$	&$1.971\pm0.154\pm0.269\pm0.139$ \\ 
		(5.0,5.5)	&$3.645\pm0.025\pm0.088\pm0.166$	&$3.189\pm0.026\pm0.081\pm0.147$	&$2.725\pm0.034\pm0.108\pm0.139$	&$2.307\pm0.082\pm0.158\pm0.119$	&			-				 \\ 
		(5.5,6.0)	&$2.611\pm0.021\pm0.074\pm0.114$	&$2.296\pm0.024\pm0.068\pm0.111$	&$1.847\pm0.030\pm0.077\pm0.093$	&$1.510\pm0.094\pm0.152\pm0.080$	&			-				 \\ 
		(6.0,7.0)	&$1.656\pm0.013\pm0.052\pm0.073$	&$1.372\pm0.014\pm0.056\pm0.064$	&$0.983\pm0.023\pm0.049\pm0.049$	&$0.717\pm0.073\pm0.180\pm0.039$	&			-				 \\ 
		(7.0,8.0)	&$0.924\pm0.010\pm0.033\pm0.042$	&$0.746\pm0.012\pm0.034\pm0.035$	&$0.493\pm0.021\pm0.044\pm0.030$	&			-					&			-				 \\ 
		(8.0,9.0)	&$0.533\pm0.009\pm0.018\pm0.025$	&$0.405\pm0.014\pm0.025\pm0.021$	&$0.202\pm0.028\pm0.038\pm0.011$	&			-					&			-				 \\ 
		(9.0,10.0)	&$0.287\pm0.008\pm0.010\pm0.014$	&$0.243\pm0.013\pm0.018\pm0.013$	&			-					&			-					&			-				 \\ 
		(10.0,11.0)	&$0.161\pm0.007\pm0.011\pm0.008$	&$0.129\pm0.015\pm0.024\pm0.008$	&			-					&			-					&			-				 \\ 
		(11.0,12.0)	&$0.110\pm0.008\pm0.009\pm0.007$	&			-					&			-					&			-					&			-				 \\ 
		(12.0,13.0)	&$0.079\pm0.007\pm0.011\pm0.005$	&			-					&			-					&			-					&			-				 \\ 
		(13.0,15.0)	&$0.051\pm0.009\pm0.017\pm0.004$	&			-					&			-					&			-					&			-				 \\ 
		(15.0,30.0)	&			-					&			-					&			-					&			-					&			-				 \\ 
 		\hline 
	\end{tabular}\label{tab:cross_2D1} 
	}
\end{sidewaystable}
\begin{sidewaystable}[htbp] 
	\caption{Double-differential cross-sections for prompt \Dz mesons in intervals of
    \pt and $y^*$ in backward rapidity regions.
    The first uncertainty is statistical, the second is the component
    of the systematic uncertainty that is uncorrelated across intervals and the third is the correlated component.}
	\centering 
	\scalebox{0.7}{ 
	\begin{tabular}{cccccc} 
		\hline 
		\multicolumn{6}{c}{$\deriv^2 \sigma/(\deriv \pt \deriv y)~[\mathrm{mb}/(\gevc)]$} \\ 
		$\pt~[\gevc]\backslash y^*$	&(-2.75, -2.50)	&(-3.00, -2.75)	&(-3.25, -3.00)	&(-3.50, -3.25)	&(-3.75, -3.50) \\ 
		\hline 
		(0.0,1.0)	&$26.423\pm0.077\pm1.135\pm2.046$	&$30.978\pm0.110\pm0.658\pm2.272$	&$31.059\pm0.086\pm0.746\pm1.764$	&$32.488\pm0.079\pm0.575\pm1.859$	&$30.706\pm0.075\pm0.555\pm1.731$ \\ 
		(1.0,1.5)	&$50.258\pm0.329\pm1.766\pm3.570$	&$55.180\pm0.193\pm1.433\pm3.295$	&$56.266\pm0.152\pm1.055\pm3.169$	&$55.760\pm0.139\pm0.959\pm2.935$	&$53.809\pm0.143\pm0.971\pm2.852$ \\ 
		(1.5,2.0)	&$51.254\pm0.277\pm1.579\pm3.222$	&$51.164\pm0.162\pm0.940\pm2.918$	&$52.039\pm0.133\pm1.058\pm2.768$	&$50.232\pm0.125\pm0.751\pm2.968$	&$47.612\pm0.121\pm1.373\pm2.535$ \\ 
		(2.0,2.5)	&$41.615\pm0.211\pm1.109\pm2.506$	&$42.476\pm0.126\pm1.047\pm2.332$	&$41.267\pm0.099\pm0.490\pm2.482$	&$39.322\pm0.092\pm0.618\pm2.022$	&$36.252\pm0.094\pm1.045\pm1.873$ \\ 
		(2.5,3.0)	&$27.597\pm0.140\pm0.693\pm1.672$	&$29.397\pm0.089\pm0.468\pm1.603$	&$29.188\pm0.150\pm0.490\pm1.570$	&$27.301\pm0.062\pm0.725\pm1.417$	&$24.672\pm0.080\pm0.394\pm1.389$ \\ 
		(3.0,3.5)	&$20.272\pm0.106\pm0.478\pm1.148$	&$20.175\pm0.079\pm0.294\pm1.079$	&$20.078\pm0.055\pm0.590\pm1.055$	&$18.466\pm0.049\pm0.464\pm0.946$	&$16.460\pm0.056\pm0.292\pm0.805$ \\ 
		(3.5,4.0)	&$13.875\pm0.078\pm0.362\pm0.737$	&$13.732\pm0.046\pm0.269\pm0.727$	&$13.085\pm0.079\pm0.394\pm0.678$	&$12.130\pm0.046\pm0.221\pm0.637$	&$10.876\pm0.041\pm0.303\pm0.544$ \\ 
		(4.0,4.5)	&$9.523\pm0.109\pm0.258\pm0.517$	&$9.049\pm0.040\pm0.232\pm0.515$	&$8.776\pm0.039\pm0.182\pm0.488$	&$8.053\pm0.034\pm0.160\pm0.407$	&$6.978\pm0.032\pm0.145\pm0.359$ \\ 
		(4.5,5.0)	&$6.647\pm0.044\pm0.182\pm0.363$	&$6.474\pm0.043\pm0.190\pm0.325$	&$6.065\pm0.028\pm0.132\pm0.294$	&$5.284\pm0.026\pm0.124\pm0.269$	&$4.464\pm0.022\pm0.105\pm0.216$ \\ 
		(5.0,5.5)	&$4.622\pm0.035\pm0.153\pm0.247$	&$4.353\pm0.032\pm0.106\pm0.228$	&$3.888\pm0.030\pm0.093\pm0.202$	&$3.662\pm0.023\pm0.089\pm0.184$	&$3.006\pm0.019\pm0.091\pm0.150$ \\ 
		(5.5,6.0)	&$3.160\pm0.041\pm0.115\pm0.172$	&$2.878\pm0.028\pm0.089\pm0.158$	&$2.741\pm0.024\pm0.086\pm0.133$	&$2.454\pm0.017\pm0.068\pm0.125$	&$2.209\pm0.016\pm0.064\pm0.114$ \\ 
		(6.0,7.0)	&$1.893\pm0.023\pm0.098\pm0.105$	&$1.798\pm0.015\pm0.044\pm0.094$	&$1.669\pm0.012\pm0.047\pm0.084$	&$1.478\pm0.010\pm0.037\pm0.087$	&$1.225\pm0.009\pm0.039\pm0.066$ \\ 
		(7.0,8.0)	&$1.021\pm0.012\pm0.034\pm0.055$	&$1.055\pm0.007\pm0.030\pm0.054$	&$0.807\pm0.009\pm0.027\pm0.044$	&$0.824\pm0.007\pm0.026\pm0.054$	&$0.646\pm0.006\pm0.021\pm0.039$ \\ 
		(8.0,9.0)	&$0.632\pm0.014\pm0.014\pm0.033$	&$0.562\pm0.006\pm0.020\pm0.029$	&$0.467\pm0.006\pm0.011\pm0.029$	&$0.429\pm0.005\pm0.008\pm0.030$	&$0.355\pm0.005\pm0.010\pm0.022$ \\ 
		(9.0,10.0)	&$0.377\pm0.006\pm0.010\pm0.019$	&$0.312\pm0.006\pm0.008\pm0.017$	&$0.266\pm0.005\pm0.007\pm0.017$	&$0.256\pm0.004\pm0.007\pm0.020$	&$0.194\pm0.004\pm0.006\pm0.013$ \\ 
		(10.0,11.0)	&$0.241\pm0.005\pm0.008\pm0.013$	&$0.184\pm0.003\pm0.006\pm0.011$	&$0.151\pm0.003\pm0.004\pm0.011$	&$0.157\pm0.003\pm0.005\pm0.014$	&$0.117\pm0.003\pm0.008\pm0.009$ \\ 
		(11.0,12.0)	&$0.155\pm0.009\pm0.007\pm0.008$	&$0.118\pm0.006\pm0.006\pm0.008$	&$0.102\pm0.003\pm0.004\pm0.010$	&$0.092\pm0.003\pm0.004\pm0.011$	&$0.080\pm0.003\pm0.005\pm0.006$ \\ 
		(12.0,13.0)	&$0.093\pm0.003\pm0.007\pm0.006$	&$0.080\pm0.003\pm0.004\pm0.010$	&$0.067\pm0.002\pm0.004\pm0.008$	&$0.067\pm0.002\pm0.003\pm0.008$	&$0.041\pm0.002\pm0.004\pm0.004$ \\ 
		(13.0,15.0)	&$0.056\pm0.002\pm0.002\pm0.005$	&$0.046\pm0.002\pm0.003\pm0.012$	&$0.033\pm0.001\pm0.001\pm0.005$	&$0.033\pm0.002\pm0.002\pm0.005$	&$0.024\pm0.002\pm0.003\pm0.002$ \\ 
        (15.0,30.0)	&$(8.6\pm0.3\pm0.6\pm2.4)\times 10^{-3}$    &$(5.5\pm0.2\pm0.4\pm1.5)\times 10^{-3}$    &$(4.6\pm0.2\pm0.3\pm0.9)\times 10^{-3}$    &$(6.1\pm0.4\pm0.6\pm1.1)\times 10^{-3}$    &           -                \\ 
		\hline 
		$\pt~[\gevc]\backslash y^*$	&(-4.00, -3.75)	&(-4.25, -4.00)	&(-4.50, -4.25)	&(-4.75, -4.50)	&(-5.00, -4.75) \\ 
		\hline 
		(0.0,1.0)	&$30.648\pm0.072\pm0.728\pm1.785$	&$30.899\pm0.087\pm0.480\pm1.646$	&$30.195\pm0.102\pm1.591\pm1.724$	&$27.540\pm0.167\pm2.236\pm1.559$	&$27.064\pm0.176\pm1.855\pm1.897$ \\ 
		(1.0,1.5)	&$50.516\pm0.149\pm1.295\pm2.628$	&$49.920\pm0.154\pm1.026\pm2.753$	&$45.676\pm0.273\pm1.648\pm2.581$	&$39.963\pm0.244\pm1.630\pm2.396$	&$37.309\pm0.360\pm2.581\pm2.781$ \\ 
		(1.5,2.0)	&$46.320\pm0.125\pm1.204\pm2.395$	&$42.339\pm0.125\pm1.000\pm2.246$	&$36.955\pm0.155\pm1.076\pm2.017$	&$33.341\pm0.208\pm0.936\pm2.091$	&$29.194\pm0.518\pm1.784\pm2.086$ \\ 
		(2.0,2.5)	&$33.509\pm0.107\pm0.558\pm1.718$	&$30.329\pm0.102\pm0.515\pm1.557$	&$27.392\pm0.119\pm0.550\pm1.415$	&$22.942\pm0.159\pm0.702\pm1.405$	&$19.056\pm0.326\pm1.006\pm1.649$ \\ 
		(2.5,3.0)	&$21.950\pm0.069\pm0.390\pm1.114$	&$20.166\pm0.078\pm0.350\pm1.076$	&$17.099\pm0.093\pm0.526\pm1.266$	&$13.958\pm0.096\pm0.359\pm0.918$	&$10.073\pm0.161\pm0.579\pm0.860$ \\ 
		(3.0,3.5)	&$14.270\pm0.055\pm0.262\pm0.720$	&$12.483\pm0.054\pm0.235\pm0.681$	&$10.923\pm0.060\pm0.233\pm0.656$	&$8.404\pm0.075\pm0.256\pm0.580$	&$5.923\pm0.129\pm0.373\pm0.566$ \\ 
		(3.5,4.0)	&$9.224\pm0.037\pm0.187\pm0.462$	&$8.029\pm0.040\pm0.239\pm0.453$	&$6.325\pm0.044\pm0.186\pm0.394$	&$4.711\pm0.058\pm0.172\pm0.368$	&$3.459\pm0.169\pm0.395\pm0.339$ \\ 
		(4.0,4.5)	&$5.943\pm0.029\pm0.119\pm0.306$	&$4.791\pm0.028\pm0.114\pm0.295$	&$4.084\pm0.035\pm0.116\pm0.272$	&$2.899\pm0.050\pm0.177\pm0.231$	&$1.790\pm0.116\pm0.282\pm0.188$ \\ 
		(4.5,5.0)	&$3.863\pm0.026\pm0.098\pm0.208$	&$3.150\pm0.024\pm0.094\pm0.196$	&$2.502\pm0.030\pm0.088\pm0.165$	&$1.621\pm0.048\pm0.135\pm0.142$	&			-				 \\ 
		(5.0,5.5)	&$2.708\pm0.019\pm0.076\pm0.154$	&$1.993\pm0.019\pm0.070\pm0.131$	&$1.598\pm0.025\pm0.082\pm0.122$	&$1.001\pm0.072\pm0.117\pm0.093$	&			-				 \\ 
		(5.5,6.0)	&$1.722\pm0.015\pm0.053\pm0.098$	&$1.389\pm0.016\pm0.052\pm0.087$	&$0.895\pm0.021\pm0.058\pm0.066$	&$0.458\pm0.049\pm0.086\pm0.040$	&			-				 \\ 
		(6.0,7.0)	&$0.945\pm0.008\pm0.031\pm0.063$	&$0.736\pm0.010\pm0.027\pm0.054$	&$0.550\pm0.023\pm0.048\pm0.046$	&			-					&			-				 \\ 
		(7.0,8.0)	&$0.489\pm0.007\pm0.020\pm0.032$	&$0.371\pm0.010\pm0.026\pm0.028$	&$0.241\pm0.023\pm0.053\pm0.024$	&			-					&			-				 \\ 
		(8.0,9.0)	&$0.258\pm0.005\pm0.009\pm0.018$	&$0.153\pm0.010\pm0.011\pm0.012$	&			-					&			-					&			-				 \\ 
		(9.0,10.0)	&$0.128\pm0.006\pm0.007\pm0.010$	&$0.083\pm0.013\pm0.014\pm0.007$	&			-					&			-					&			-				 \\ 
		(10.0,11.0)	&$0.069\pm0.004\pm0.006\pm0.006$	&			-					&			-					&			-					&			-				 \\ 
		(11.0,12.0)	&$0.033\pm0.004\pm0.005\pm0.003$	&			-					&			-					&			-					&			-				 \\ 
		(12.0,13.0)	&			-					&			-					&			-					&			-					&			-				 \\ 
		(13.0,15.0)	&			-					&			-					&			-					&			-					&			-				 \\ 
		(15.0,30.0)	&			-					&			-					&			-					&			-					&			-				 \\ 
 		\hline 
	\end{tabular}\label{tab:cross_2D2} 
	}
\end{sidewaystable}
\begin{table}[htbp] 
	\caption{Nuclear modification factor $R_{p\mathrm{Pb}}$ for prompt \Dz mesons
    in intervals of \pt and $y^*$ for $\pt<10\gevc$.
    The first uncertainty is statistical and the second is the systematic.} 
	\centering 
	\scalebox{0.65}{ 
	\begin{tabular}{cccccc} 
		\hline 
		\multicolumn{6}{c}{$R_{p\mathrm{Pb}}$} \\ 
        $\pt~[\gevc]\backslash y^*$	&(2.5, 4.0) &(2.0, 2.5)	&(2.5, 3.0)	&(3.0, 3.5)	&(3.5, 4.0) \\ 
		\hline 
		(0.0,1.0)	&$0.546\pm0.002\pm0.033$   &$0.485\pm0.001\pm0.041$	&$0.525\pm0.001\pm0.032$	&$0.556\pm0.002\pm0.036$	&$0.561\pm0.005\pm0.039$ \\
		(1.0,2.0)	&$0.596\pm0.002\pm0.034$   &$0.557\pm0.001\pm0.037$	&$0.591\pm0.003\pm0.034$	&$0.611\pm0.002\pm0.036$	&$0.585\pm0.003\pm0.038$ \\
		(2.0,3.0)	&$0.637\pm0.001\pm0.034$   &$0.648\pm0.001\pm0.036$	&$0.637\pm0.001\pm0.034$	&$0.648\pm0.001\pm0.035$	&$0.624\pm0.003\pm0.037$ \\
		(3.0,4.0)	&$0.671\pm0.001\pm0.036$   &$0.679\pm0.001\pm0.038$	&$0.676\pm0.002\pm0.035$	&$0.673\pm0.002\pm0.036$	&$0.659\pm0.004\pm0.044$ \\
		(4.0,5.0)	&$0.706\pm0.002\pm0.040$   &$0.697\pm0.002\pm0.042$	&$0.719\pm0.002\pm0.039$	&$0.710\pm0.003\pm0.041$	&$0.681\pm0.007\pm0.048$ \\
		(5.0,6.0)	&$0.719\pm0.005\pm0.048$   &$0.718\pm0.003\pm0.056$	&$0.722\pm0.002\pm0.047$	&$0.737\pm0.004\pm0.047$	&$0.688\pm0.019\pm0.064$ \\
		(6.0,7.0)	&$0.710\pm0.014\pm0.067$   &$0.721\pm0.004\pm0.056$	&$0.769\pm0.004\pm0.058$	&$0.725\pm0.006\pm0.057$	&$0.568\pm0.061\pm0.169$ \\
		(7.0,8.0)	&$0.752\pm0.005\pm0.061$   &$0.777\pm0.006\pm0.067$	&$0.783\pm0.005\pm0.061$	&$0.709\pm0.010\pm0.067$	&		-			 \\ 
		(8.0,9.0)	&$0.768\pm0.011\pm0.073$   &$0.717\pm0.006\pm0.084$	&$0.832\pm0.008\pm0.074$	&$0.683\pm0.023\pm0.078$	&		-			 \\ 
		(9.0,10.0)	&$0.784\pm0.018\pm0.111$   &$0.687\pm0.007\pm0.070$	&$0.764\pm0.011\pm0.086$	&$0.814\pm0.043\pm0.160$	&		-			 \\ 
		\hline 
	\end{tabular}\label{tab:RpA_loose1} 
	}
\end{table}
\begin{table}[htbp] 
	\caption{Nuclear modification factor $R_{p\mathrm{Pb}}$ for prompt \Dz mesons
    in intervals of \pt and $y^*$ for $\pt<10\gevc$.
    The first uncertainty is statistical and the second is systematic.} 
	\centering 
	\scalebox{0.65}{ 
	\begin{tabular}{cccccc} 
		\hline 
		\multicolumn{6}{c}{$R_{p\mathrm{Pb}}$} \\ 
        $\pt~[\gevc]\backslash y^*$	& (-4.0, -2.5) &(-3.0, -2.5)	&(-3.5, -3.0)	&(-4.0, -3.5)	&(-4.5, -4.0) \\ 
		\hline 
		(0.0,1.0)	&$0.691\pm0.001\pm0.049$   &$0.607\pm0.002\pm0.047$	&$0.706\pm0.001\pm0.049$	&$0.781\pm0.001\pm0.054$	&$0.959\pm0.002\pm0.080$ \\ 
		(1.0,2.0)	&$0.803\pm0.001\pm0.053$   &$0.718\pm0.001\pm0.052$	&$0.824\pm0.001\pm0.054$	&$0.891\pm0.001\pm0.057$	&$0.994\pm0.002\pm0.068$ \\ 
		(2.0,3.0)	&$0.891\pm0.001\pm0.056$   &$0.804\pm0.002\pm0.056$	&$0.940\pm0.002\pm0.056$	&$0.960\pm0.001\pm0.058$	&$1.087\pm0.002\pm0.075$ \\ 
		(3.0,4.0)	&$0.917\pm0.001\pm0.053$   &$0.841\pm0.003\pm0.051$	&$0.982\pm0.002\pm0.056$	&$0.952\pm0.002\pm0.057$	&$1.134\pm0.003\pm0.099$ \\ 
		(4.0,5.0)	&$0.916\pm0.003\pm0.056$   &$0.860\pm0.008\pm0.054$	&$0.962\pm0.002\pm0.058$	&$0.948\pm0.003\pm0.062$	&$1.031\pm0.004\pm0.139$ \\ 
		(5.0,6.0)	&$0.894\pm0.002\pm0.062$   &$0.832\pm0.005\pm0.061$	&$0.929\pm0.003\pm0.060$	&$0.956\pm0.004\pm0.072$	&$1.493\pm0.010\pm0.403$ \\ 
		(6.0,7.0)	&$0.884\pm0.003\pm0.074$   &$0.819\pm0.005\pm0.064$	&$0.911\pm0.005\pm0.073$	&$0.974\pm0.005\pm0.111$	&		-			 \\ 
		(7.0,8.0)	&$0.898\pm0.004\pm0.086$   &$0.827\pm0.006\pm0.067$	&$0.881\pm0.006\pm0.084$	&$1.101\pm0.009\pm0.158$	&		-			 \\ 
		(8.0,9.0)	&$0.918\pm0.006\pm0.118$   &$0.861\pm0.010\pm0.079$	&$0.857\pm0.008\pm0.091$	&$1.197\pm0.014\pm0.352$	&		-			 \\ 
		(9.0,10.0)	&$0.867\pm0.007\pm0.123$   &$0.805\pm0.010\pm0.091$	&$0.964\pm0.011\pm0.181$	&		-				&		-			 \\ 
		\hline 
	\end{tabular}\label{tab:RpA_loose2} 
	}
\end{table}
\begin{table}[htbp] 
	\caption{Forward-backward production ratio $R_\mathrm{FB}$ for prompt \Dz mesons as a function of \pt,
    integrated over $2.5<|y^*|<4.0$.
    The first uncertainty is statistical,
    the second is the component of the systematic uncertainty that is uncorrelated
    across intervals and the third is the correlated component.}
	\centering 
	\begin{tabular}{cc} 
		\hline 
		$\pt~[\gevc]$ & $R_\mathrm{FB}$ \\ 
		\hline 
		(0.0,1.0)	&$0.808\pm0.009\pm0.019\pm0.051$ \\ 
		(1.0,1.5)	&$0.766\pm0.002\pm0.014\pm0.044$ \\ 
		(1.5,2.0)	&$0.725\pm0.002\pm0.011\pm0.041$ \\ 
		(2.0,2.5)	&$0.708\pm0.002\pm0.011\pm0.039$ \\ 
		(2.5,3.0)	&$0.732\pm0.002\pm0.011\pm0.040$ \\ 
		(3.0,3.5)	&$0.732\pm0.002\pm0.011\pm0.039$ \\ 
		(3.5,4.0)	&$0.745\pm0.003\pm0.013\pm0.040$ \\ 
		(4.0,4.5)	&$0.760\pm0.004\pm0.012\pm0.042$ \\ 
		(4.5,5.0)	&$0.760\pm0.006\pm0.015\pm0.040$ \\ 
		(5.0,5.5)	&$0.811\pm0.006\pm0.016\pm0.044$ \\ 
		(5.5,6.0)	&$0.829\pm0.009\pm0.020\pm0.045$ \\ 
		(6.0,7.0)	&$0.815\pm0.010\pm0.029\pm0.045$ \\ 
		(7.0,8.0)	&$0.868\pm0.008\pm0.024\pm0.051$ \\ 
		(8.0,9.0)	&$0.835\pm0.017\pm0.027\pm0.050$ \\ 
		(9.0,10.0)	&$0.926\pm0.019\pm0.028\pm0.055$ \\ 
		(10.0,11.0)	&$0.896\pm0.032\pm0.050\pm0.060$ \\ 
		(11.0,12.0)	&$0.925\pm0.048\pm0.051\pm0.063$ \\ 
		(12.0,13.0)	&$1.096\pm0.056\pm0.088\pm0.106$ \\ 
		(13.0,15.0)	&$1.078\pm0.099\pm0.178\pm0.181$ \\ 
		\hline 
	\end{tabular}\label{tab:RFB_pt} 
\end{table}
\clearpage

%% file: Authorship_LHCb-PAPER-2022-007.tex
\centerline
{\large\bf LHCb collaboration}
\begin
{flushleft}
\small
R.~Aaij$^{32}$\lhcborcid{0000-0003-0533-1952},
A.S.W.~Abdelmotteleb$^{50}$\lhcborcid{0000-0001-7905-0542},
C.~Abellan~Beteta$^{44}$,
F.~Abudin{\'e}n$^{50}$\lhcborcid{0000-0002-6737-3528},
T.~Ackernley$^{54}$\lhcborcid{0000-0002-5951-3498},
B.~Adeva$^{40}$\lhcborcid{0000-0001-9756-3712},
M.~Adinolfi$^{48}$\lhcborcid{0000-0002-1326-1264},
H.~Afsharnia$^{9}$,
C.~Agapopoulou$^{13}$\lhcborcid{0000-0002-2368-0147},
C.A.~Aidala$^{76}$\lhcborcid{0000-0001-9540-4988}
% ...

{\footnotesize \it

% ...
$^{9}$Universit{\'e} Clermont Auvergne, CNRS/IN2P3, LPC, Clermont-Ferrand, France\\
% ...
$^{13}$LPNHE, Sorbonne Universit{\'e}, Paris Diderot Sorbonne Paris Cit{\'e}, CNRS/IN2P3, Paris, France\\
% ...
$^{32}$Nikhef National Institute for Subatomic Physics, Amsterdam, Netherlands\\
% ...
$^{40}$Instituto Galego de F{\'\i}sica de Altas Enerx{\'\i}as (IGFAE), Universidade de Santiago de Compostela, Santiago de Compostela, Spain\\
% ...
$^{44}$Physik-Institut, Universit{\"a}t Z{\"u}rich, Z{\"u}rich, Switzerland\\
% ...
$^{48}$H.H. Wills Physics Laboratory, University of Bristol, Bristol, United Kingdom\\
% ...
$^{50}$Department of Physics, University of Warwick, Coventry, United Kingdom\\
% ...
$^{54}$Oliver Lodge Laboratory, University of Liverpool, Liverpool, United Kingdom\\
% ...
$^{62}$Syracuse University, Syracuse, NY, United States\\
% ...
$^{76}$University of Michigan, Ann Arbor, MI, United States, associated to $^{62}$\\
\bigskip
% ...
}
\end{flushleft}

%% file: main.bbl
\begin{mcitethebibliography}{10}
                    \mciteSetBstSublistMode{n}
                    \mciteSetBstMaxWidthForm{subitem}{\alph{mcitesubitemcount})}
                    \mciteSetBstSublistLabelBeginEnd{\mcitemaxwidthsubitemform\space}
                    {\relax}{\relax}

                \bibitem{Yagi:2005yb}
                    K.~Yagi, T.~Hatsuda, and Y.~Miake, {\em Quark-gluon plasma: From big bang to
                    little bang}, vol.~23, Cambridge University Press, 2005\relax
                    \mciteBstWouldAddEndPuncttrue
                    \mciteSetBstMidEndSepPunct{\mcitedefaultmidpunct}
                    {\mcitedefaultendpunct}{\mcitedefaultseppunct}\relax
                    \EndOfBibitem
                \bibitem{Eskola:2009uj}
                    K.~J. Eskola, H.~Paukkunen, and C.~A. Salgado,
                    \ifthenelse{\boolean{articletitles}}{\emph{{EPS09: A new generation of NLO
                    and LO nuclear parton distribution functions}},
                    }{}\href{https://doi.org/10.1088/1126-6708/2009/04/065}{JHEP \textbf{04}
                    (2009) 065}, \href{http://arxiv.org/abs/0902.4154}{{\normalfont\ttfamily
                    arXiv:0902.4154}}\relax
                    \mciteBstWouldAddEndPuncttrue
                    \mciteSetBstMidEndSepPunct{\mcitedefaultmidpunct}
                    {\mcitedefaultendpunct}{\mcitedefaultseppunct}\relax
                    \EndOfBibitem
                \bibitem{deFlorian:2003qf}
                    D.~de~Florian and R.~Sassot,
                    \ifthenelse{\boolean{articletitles}}{\emph{{Nuclear parton distributions at
                    next-to-leading order}},
                    }{}\href{https://doi.org/10.1103/PhysRevD.69.074028}{Phys.\ Rev.\
                    \textbf{D69} (2004) 074028},
                    \href{http://arxiv.org/abs/hep-ph/0311227}{{\normalfont\ttfamily
                    arXiv:hep-ph/0311227}}\relax
                    \mciteBstWouldAddEndPuncttrue
                    \mciteSetBstMidEndSepPunct{\mcitedefaultmidpunct}
                    {\mcitedefaultendpunct}{\mcitedefaultseppunct}\relax
                    \EndOfBibitem
                \bibitem{Hirai:2007sx}
                    M.~Hirai, S.~Kumano, and T.-H. Nagai,
                    \ifthenelse{\boolean{articletitles}}{\emph{{Determination of nuclear parton
                    distribution functions and their uncertainties in next-to-leading order}},
                    }{}\href{https://doi.org/10.1103/PhysRevC.76.065207}{Phys.\ Rev.\
                    \textbf{C76} (2007) 065207},
                    \href{http://arxiv.org/abs/0709.3038}{{\normalfont\ttfamily
                    arXiv:0709.3038}}\relax
                    \mciteBstWouldAddEndPuncttrue
                    \mciteSetBstMidEndSepPunct{\mcitedefaultmidpunct}
                    {\mcitedefaultendpunct}{\mcitedefaultseppunct}\relax
                    \EndOfBibitem
                \bibitem{Armesto:2006ph}
                    N.~Armesto, \ifthenelse{\boolean{articletitles}}{\emph{{Nuclear shadowing}},
                    }{}\href{https://doi.org/10.1088/0954-3899/32/11/R01}{J.\ Phys.\
                    \textbf{G32} (2006) R367},
                    \href{http://arxiv.org/abs/hep-ph/0604108}{{\normalfont\ttfamily
                    arXiv:hep-ph/0604108}}\relax
                    \mciteBstWouldAddEndPuncttrue
                    \mciteSetBstMidEndSepPunct{\mcitedefaultmidpunct}
                    {\mcitedefaultendpunct}{\mcitedefaultseppunct}\relax
                    \EndOfBibitem
                \bibitem{Gelis:2010nm}
                    F.~Gelis, E.~Iancu, J.~Jalilian-Marian, and R.~Venugopalan,
                    \ifthenelse{\boolean{articletitles}}{\emph{{The color glass condensate}},
                    }{}\href{https://doi.org/10.1146/annurev.nucl.010909.083629}{Ann.\ Rev.\
                    Nucl.\ Part.\ Sci.\  \textbf{60} (2010) 463},
                    \href{http://arxiv.org/abs/1002.0333}{{\normalfont\ttfamily
                    arXiv:1002.0333}}\relax
                    \mciteBstWouldAddEndPuncttrue
                    \mciteSetBstMidEndSepPunct{\mcitedefaultmidpunct}
                    {\mcitedefaultendpunct}{\mcitedefaultseppunct}\relax
                    \EndOfBibitem
                \bibitem{Vitev:2007ve}
                    I.~Vitev, \ifthenelse{\boolean{articletitles}}{\emph{{Non-Abelian energy loss
                    in cold nuclear matter}},
                    }{}\href{https://doi.org/10.1103/PhysRevC.75.064906}{Phys.\ Rev.\
                    \textbf{C75} (2007) 064906},
                    \href{http://arxiv.org/abs/hep-ph/0703002}{{\normalfont\ttfamily
                    arXiv:hep-ph/0703002}}\relax
                    \mciteBstWouldAddEndPuncttrue
                    \mciteSetBstMidEndSepPunct{\mcitedefaultmidpunct}
                    {\mcitedefaultendpunct}{\mcitedefaultseppunct}\relax
                    \EndOfBibitem
                \bibitem{Kang:2014hha}
                    Z.-B. Kang {\em et~al.}, \ifthenelse{\boolean{articletitles}}{\emph{{Multiple
                    scattering effects on heavy meson production in p+A collisions at backward
                    rapidity}}, }{}\href{https://doi.org/10.1016/j.physletb.2014.11.024}{Phys.\
                    Lett.\ \textbf{B740} (2015) 23},
                    \href{http://arxiv.org/abs/1409.2494}{{\normalfont\ttfamily
                    arXiv:1409.2494}}\relax
                    \mciteBstWouldAddEndPuncttrue
                    \mciteSetBstMidEndSepPunct{\mcitedefaultmidpunct}
                    {\mcitedefaultendpunct}{\mcitedefaultseppunct}\relax
                    \EndOfBibitem
                \bibitem{Arleo:2021bpv}
                    F.~Arleo, G.~Jackson, and S.~Peign\'e,
                    \ifthenelse{\boolean{articletitles}}{\emph{{Impact of fully coherent energy
                    loss on heavy meson production in pA collisions}},
                    }{}\href{https://doi.org/10.1007/JHEP01(2022)164}{JHEP \textbf{01} (2022)
                    164}, \href{http://arxiv.org/abs/2107.05871}{{\normalfont\ttfamily
                    arXiv:2107.05871}}\relax
                    \mciteBstWouldAddEndPuncttrue
                    \mciteSetBstMidEndSepPunct{\mcitedefaultmidpunct}
                    {\mcitedefaultendpunct}{\mcitedefaultseppunct}\relax
                    \EndOfBibitem
                \bibitem{Zhang:2019dth}
                    C.~Zhang {\em et~al.}, \ifthenelse{\boolean{articletitles}}{\emph{{Elliptic
                    flow of heavy quarkonia in $pA$ collisions}},
                    }{}\href{https://doi.org/10.1103/PhysRevLett.122.172302}{Phys.\ Rev.\ Lett.\
                    \textbf{122} (2019) 172302},
                    \href{http://arxiv.org/abs/1901.10320}{{\normalfont\ttfamily
                    arXiv:1901.10320}}\relax
                    \mciteBstWouldAddEndPuncttrue
                    \mciteSetBstMidEndSepPunct{\mcitedefaultmidpunct}
                    {\mcitedefaultendpunct}{\mcitedefaultseppunct}\relax
                    \EndOfBibitem
                \bibitem{Zhao:2020wcd}
                    W.~Zhao {\em et~al.}, \ifthenelse{\boolean{articletitles}}{\emph{{Probing the
                    Partonic Degrees of Freedom in High-Multiplicity $p-Pb$ collisions at $\sqrt
                    {s_{NN}}$ = 5.02 TeV}},
                    }{}\href{https://doi.org/10.1103/PhysRevLett.125.072301}{Phys.\ Rev.\ Lett.\
                    \textbf{125} (2020) 072301},
                    \href{http://arxiv.org/abs/1911.00826}{{\normalfont\ttfamily
                    arXiv:1911.00826}}\relax
                    \mciteBstWouldAddEndPuncttrue
                    \mciteSetBstMidEndSepPunct{\mcitedefaultmidpunct}
                    {\mcitedefaultendpunct}{\mcitedefaultseppunct}\relax
                    \EndOfBibitem
                \bibitem{Sirunyan:2018toe}
                    CMS collaboration, A.~M. Sirunyan {\em et~al.},
                    \ifthenelse{\boolean{articletitles}}{\emph{{Elliptic flow of charm and
                    strange hadrons in high-multiplicity pPb collisions at $\sqsnn = $8.16\tev}},
                    }{}\href{https://doi.org/10.1103/PhysRevLett.121.082301}{Phys.\ Rev.\ Lett.\
                    \textbf{121} (2018) 082301},
                    \href{http://arxiv.org/abs/1804.09767}{{\normalfont\ttfamily
                    arXiv:1804.09767}}\relax
                    \mciteBstWouldAddEndPuncttrue
                    \mciteSetBstMidEndSepPunct{\mcitedefaultmidpunct}
                    {\mcitedefaultendpunct}{\mcitedefaultseppunct}\relax
                    \EndOfBibitem
                \bibitem{Sirunyan:2018kiz}
                    CMS collaboration, A.~M. Sirunyan {\em et~al.},
                    \ifthenelse{\boolean{articletitles}}{\emph{{Observation of prompt J/$\psi$
                    meson elliptic flow in high-multiplicity pPb collisions at $\sqsnn =
                    $8.16\tev}}, }{}\href{https://doi.org/10.1016/j.physletb.2019.02.018}{Phys.\
                    Lett.\  \textbf{B791} (2019) 172},
                    \href{http://arxiv.org/abs/1810.01473}{{\normalfont\ttfamily
                    arXiv:1810.01473}}\relax
                    \mciteBstWouldAddEndPuncttrue
                    \mciteSetBstMidEndSepPunct{\mcitedefaultmidpunct}
                    {\mcitedefaultendpunct}{\mcitedefaultseppunct}\relax
                    \EndOfBibitem
                \bibitem{LHCb-PAPER-2017-015}
                    LHCb collaboration, R.~Aaij {\em et~al.},
                    \ifthenelse{\boolean{articletitles}}{\emph{{Study of prompt \Dz meson
                    production in \proton{}Pb collisions at $\sqsnn = $5\tev}},
                    }{}\href{https://doi.org/10.1007/JHEP10(2017)090}{JHEP \textbf{10} (2017)
                    090}, \href{http://arxiv.org/abs/1707.02750}{{\normalfont\ttfamily
                    arXiv:1707.02750}}\relax
                    \mciteBstWouldAddEndPuncttrue
                    \mciteSetBstMidEndSepPunct{\mcitedefaultmidpunct}
                    {\mcitedefaultendpunct}{\mcitedefaultseppunct}\relax
                    \EndOfBibitem
                \bibitem{LHCb-PAPER-2018-021}
                    LHCb collaboration, R.~Aaij {\em et~al.},
                    \ifthenelse{\boolean{articletitles}}{\emph{{Prompt \Lc production in
                    \proton{}Pb collisions at $\sqsnn = $5.02\tev}},
                    }{}\href{https://doi.org/10.1007/JHEP02(2019)102}{JHEP \textbf{02} (2019)
                    102}, \href{http://arxiv.org/abs/1809.01404}{{\normalfont\ttfamily
                    arXiv:1809.01404}}\relax
                    \mciteBstWouldAddEndPuncttrue
                    \mciteSetBstMidEndSepPunct{\mcitedefaultmidpunct}
                    {\mcitedefaultendpunct}{\mcitedefaultseppunct}\relax
                    \EndOfBibitem
                \bibitem{LHCb-PAPER-2017-014}
                    LHCb collaboration, R.~Aaij {\em et~al.},
                    \ifthenelse{\boolean{articletitles}}{\emph{{Prompt and nonprompt \jpsi\
                    production and nuclear modification in \proton{}Pb collisions at $\sqsnn =
                    $8.16\tev}}, }{}\href{https://doi.org/10.1016/j.physletb.2017.09.058}{Phys.\
                    Lett.\  \textbf{B774} (2017) 159},
                    \href{http://arxiv.org/abs/1706.07122}{{\normalfont\ttfamily
                    arXiv:1706.07122}}\relax
                    \mciteBstWouldAddEndPuncttrue
                    \mciteSetBstMidEndSepPunct{\mcitedefaultmidpunct}
                    {\mcitedefaultendpunct}{\mcitedefaultseppunct}\relax
                    \EndOfBibitem
                \bibitem{LHCb-PAPER-2018-048}
                    LHCb collaboration, R.~Aaij {\em et~al.},
                    \ifthenelse{\boolean{articletitles}}{\emph{{Measurement of \Bp, \Bz and \Lb
                    production in \proton{}Pb collisions at $\sqsnn = $8.16\tev}},
                    }{}\href{https://doi.org/10.1103/PhysRevD.99.052011}{Phys.\ Rev.\
                    \textbf{D99} (2019) 052011},
                    \href{http://arxiv.org/abs/1902.05599}{{\normalfont\ttfamily
                    arXiv:1902.05599}}\relax
                    \mciteBstWouldAddEndPuncttrue
                    \mciteSetBstMidEndSepPunct{\mcitedefaultmidpunct}
                    {\mcitedefaultendpunct}{\mcitedefaultseppunct}\relax
                    \EndOfBibitem
                \bibitem{LHCb-PAPER-2018-035}
                    LHCb collaboration, R.~Aaij {\em et~al.},
                    \ifthenelse{\boolean{articletitles}}{\emph{{Study of \Upsilonres production
                    in \proton{}Pb collisions at $\sqsnn = $8.16\tev}},
                    }{}\href{https://doi.org/10.1007/JHEP11(2018)194}{JHEP \textbf{11} (2018)
                    194}, \href{http://arxiv.org/abs/1810.07655}{{\normalfont\ttfamily
                    arXiv:1810.07655}}\relax
                    \mciteBstWouldAddEndPuncttrue
                    \mciteSetBstMidEndSepPunct{\mcitedefaultmidpunct}
                    {\mcitedefaultendpunct}{\mcitedefaultseppunct}\relax
                    \EndOfBibitem
                \bibitem{ALICE:2014xjz}
                    ALICE collaboration, B.~B. Abelev {\em et~al.},
                    \ifthenelse{\boolean{articletitles}}{\emph{{Measurement of prompt $D$-meson
                    production in $p-Pb$ collisions at $\sqsnn = $5.02\tev}},
                    }{}\href{https://doi.org/10.1103/PhysRevLett.113.232301}{Phys.\ Rev.\ Lett.\
                    \textbf{113} (2014) 232301},
                    \href{http://arxiv.org/abs/1405.3452}{{\normalfont\ttfamily
                    arXiv:1405.3452}}\relax
                    \mciteBstWouldAddEndPuncttrue
                    \mciteSetBstMidEndSepPunct{\mcitedefaultmidpunct}
                    {\mcitedefaultendpunct}{\mcitedefaultseppunct}\relax
                    \EndOfBibitem
                \bibitem{ALICE:2016cpm}
                    ALICE collaboration, J.~Adam {\em et~al.},
                    \ifthenelse{\boolean{articletitles}}{\emph{{Measurement of D-meson production
                    versus multiplicity in p-Pb collisions at $\sqsnn = $5.02\tev}},
                    }{}\href{https://doi.org/10.1007/JHEP08(2016)078}{JHEP \textbf{08} (2016)
                    078}, \href{http://arxiv.org/abs/1602.07240}{{\normalfont\ttfamily
                    arXiv:1602.07240}}\relax
                    \mciteBstWouldAddEndPuncttrue
                    \mciteSetBstMidEndSepPunct{\mcitedefaultmidpunct}
                    {\mcitedefaultendpunct}{\mcitedefaultseppunct}\relax
                    \EndOfBibitem
                \bibitem{ALICE:2016yta}
                    ALICE collaboration, J.~Adam {\em et~al.},
                    \ifthenelse{\boolean{articletitles}}{\emph{{$D$-meson production in $p$-Pb
                    collisions at $\sqsnn = $5.02\tev and in pp collisions at $\sqs = $7\tev}},
                    }{}\href{https://doi.org/10.1103/PhysRevC.94.054908}{Phys.\ Rev.\
                    \textbf{C94} (2016) 054908},
                    \href{http://arxiv.org/abs/1605.07569}{{\normalfont\ttfamily
                    arXiv:1605.07569}}\relax
                    \mciteBstWouldAddEndPuncttrue
                    \mciteSetBstMidEndSepPunct{\mcitedefaultmidpunct}
                    {\mcitedefaultendpunct}{\mcitedefaultseppunct}\relax
                    \EndOfBibitem
                \bibitem{ALICE:2019fhe}
                    ALICE collaboration, S.~Acharya {\em et~al.},
                    \ifthenelse{\boolean{articletitles}}{\emph{{Measurement of prompt \Dz, \Dp,
                    \Dstarp, and \Ds production in p\textendash{}Pb collisions at $\sqsnn =
                    $5.02\tev}}, }{}\href{https://doi.org/10.1007/JHEP12(2019)092}{JHEP
                    \textbf{12} (2019) 092},
                    \href{http://arxiv.org/abs/1906.03425}{{\normalfont\ttfamily
                    arXiv:1906.03425}}\relax
                    \mciteBstWouldAddEndPuncttrue
                    \mciteSetBstMidEndSepPunct{\mcitedefaultmidpunct}
                    {\mcitedefaultendpunct}{\mcitedefaultseppunct}\relax
                    \EndOfBibitem
                \bibitem{ALICE:2020wfu}
                    ALICE collaboration, S.~Acharya {\em et~al.},
                    \ifthenelse{\boolean{articletitles}}{\emph{{\Lc production and
                    baryon-to-meson ratios in pp and p-Pb collisions at $\sqsnn = $5.02\tev at
                    the LHC}}, }{}\href{https://doi.org/10.1103/PhysRevLett.127.202301}{Phys.\
                    Rev.\ Lett.\  \textbf{127} (2021) 202301},
                    \href{http://arxiv.org/abs/2011.06078}{{\normalfont\ttfamily
                    arXiv:2011.06078}}\relax
                    \mciteBstWouldAddEndPuncttrue
                    \mciteSetBstMidEndSepPunct{\mcitedefaultmidpunct}
                    {\mcitedefaultendpunct}{\mcitedefaultseppunct}\relax
                    \EndOfBibitem
                \bibitem{ALICE:2020wla}
                    ALICE collaboration, S.~Acharya {\em et~al.},
                    \ifthenelse{\boolean{articletitles}}{\emph{{\Lc production in $pp$ and in
                    $p$-Pb collisions at $\sqsnn = $5.02\tev}},
                    }{}\href{https://doi.org/10.1103/PhysRevC.104.054905}{Phys.\ Rev.\
                    \textbf{C104} (2021) 054905},
                    \href{http://arxiv.org/abs/2011.06079}{{\normalfont\ttfamily
                    arXiv:2011.06079}}\relax
                    \mciteBstWouldAddEndPuncttrue
                    \mciteSetBstMidEndSepPunct{\mcitedefaultmidpunct}
                    {\mcitedefaultendpunct}{\mcitedefaultseppunct}\relax
                    \EndOfBibitem
                \bibitem{ALICE:2017fsl}
                    ALICE collaboration, S.~Acharya {\em et~al.},
                    \ifthenelse{\boolean{articletitles}}{\emph{{Production of muons from
                    heavy-flavour hadron decays in p-Pb collisions at $\sqsnn = $5.02\tev}},
                    }{}\href{https://doi.org/10.1016/j.physletb.2017.03.049}{Phys.\ Lett.\
                    \textbf{B770} (2017) 459},
                    \href{http://arxiv.org/abs/1702.01479}{{\normalfont\ttfamily
                    arXiv:1702.01479}}\relax
                    \mciteBstWouldAddEndPuncttrue
                    \mciteSetBstMidEndSepPunct{\mcitedefaultmidpunct}
                    {\mcitedefaultendpunct}{\mcitedefaultseppunct}\relax
                    \EndOfBibitem
                \bibitem{ALICE:2020vjy}
                    ALICE collaboration, S.~Acharya {\em et~al.},
                    \ifthenelse{\boolean{articletitles}}{\emph{{Measurement of nuclear effects on
                    $\psi(2S)$ production in p-Pb collisions at $\sqsnn = $8.16\tev}},
                    }{}\href{https://doi.org/10.1007/JHEP07(2020)237}{JHEP \textbf{07} (2020)
                    237}, \href{http://arxiv.org/abs/2003.06053}{{\normalfont\ttfamily
                    arXiv:2003.06053}}\relax
                    \mciteBstWouldAddEndPuncttrue
                    \mciteSetBstMidEndSepPunct{\mcitedefaultmidpunct}
                    {\mcitedefaultendpunct}{\mcitedefaultseppunct}\relax
                    \EndOfBibitem
                \bibitem{ALICE:2019qie}
                    ALICE collaboration, S.~Acharya {\em et~al.},
                    \ifthenelse{\boolean{articletitles}}{\emph{{$\PUpsilon$ production in p-Pb
                    collisions at $\sqsnn = $8.16\tev}},
                    }{}\href{https://doi.org/10.1016/j.physletb.2020.135486}{Phys.\ Lett.\
                    \textbf{B806} (2020) 135486},
                    \href{http://arxiv.org/abs/1910.14405}{{\normalfont\ttfamily
                    arXiv:1910.14405}}\relax
                    \mciteBstWouldAddEndPuncttrue
                    \mciteSetBstMidEndSepPunct{\mcitedefaultmidpunct}
                    {\mcitedefaultendpunct}{\mcitedefaultseppunct}\relax
                    \EndOfBibitem
                \bibitem{ALICE:2018mml}
                    ALICE collaboration, S.~Acharya {\em et~al.},
                    \ifthenelse{\boolean{articletitles}}{\emph{{Inclusive \jpsi production at
                    forward and backward rapidity in p-Pb collisions at $\sqsnn = $8.16\tev}},
                    }{}\href{https://doi.org/10.1007/JHEP07(2018)160}{JHEP \textbf{07} (2018)
                    160}, \href{http://arxiv.org/abs/1805.04381}{{\normalfont\ttfamily
                    arXiv:1805.04381}}\relax
                    \mciteBstWouldAddEndPuncttrue
                    \mciteSetBstMidEndSepPunct{\mcitedefaultmidpunct}
                    {\mcitedefaultendpunct}{\mcitedefaultseppunct}\relax
                    \EndOfBibitem
                \bibitem{ALICE:2018szk}
                    ALICE collaboration, S.~Acharya {\em et~al.},
                    \ifthenelse{\boolean{articletitles}}{\emph{{Prompt and non-prompt \jpsi
                    production and nuclear modification at mid-rapidity in p-Pb collisions at
                    $\sqsnn = $5.02\tev}},
                    }{}\href{https://doi.org/10.1140/epjc/s10052-018-5881-2}{Eur.\ Phys.\ J.\
                    \textbf{C78} (2018) 466},
                    \href{http://arxiv.org/abs/1802.00765}{{\normalfont\ttfamily
                    arXiv:1802.00765}}\relax
                    \mciteBstWouldAddEndPuncttrue
                    \mciteSetBstMidEndSepPunct{\mcitedefaultmidpunct}
                    {\mcitedefaultendpunct}{\mcitedefaultseppunct}\relax
                    \EndOfBibitem
                \bibitem{ALICE:2014ict}
                    ALICE collaboration, B.~B. Abelev {\em et~al.},
                    \ifthenelse{\boolean{articletitles}}{\emph{{Production of inclusive $\PUpsilon(1S)$ and
                    $\PUpsilon(2S)$ in p-Pb collisions at $\sqsnn = $5.02\tev}},
                    }{}\href{https://doi.org/10.1016/j.physletb.2014.11.041}{Phys.\ Lett.\
                    \textbf{B740} (2015) 105},
                    \href{http://arxiv.org/abs/1410.2234}{{\normalfont\ttfamily
                    arXiv:1410.2234}}\relax
                    \mciteBstWouldAddEndPuncttrue
                    \mciteSetBstMidEndSepPunct{\mcitedefaultmidpunct}
                    {\mcitedefaultendpunct}{\mcitedefaultseppunct}\relax
                    \EndOfBibitem
                \bibitem{ATLAS:2017prf}
                    ATLAS collaboration, M.~Aaboud {\em et~al.},
                    \ifthenelse{\boolean{articletitles}}{\emph{{Measurement of quarkonium
                    production in proton-lead and proton-proton collisions at 5.02\tev with the
                    ATLAS detector}},
                    }{}\href{https://doi.org/10.1140/epjc/s10052-018-5624-4}{Eur.\ Phys.\ J.\
                    \textbf{C78} (2018) 171},
                    \href{http://arxiv.org/abs/1709.03089}{{\normalfont\ttfamily
                    arXiv:1709.03089}}\relax
                    \mciteBstWouldAddEndPuncttrue
                    \mciteSetBstMidEndSepPunct{\mcitedefaultmidpunct}
                    {\mcitedefaultendpunct}{\mcitedefaultseppunct}\relax
                    \EndOfBibitem
                \bibitem{CMS:2022wfi}
                    CMS collaboration, A.~Tumasyan {\em et~al.},
                    \ifthenelse{\boolean{articletitles}}{\emph{{Nuclear modification of
                    $\Upsilon$ states in pPb collisions at $\sqrt{s_\mathrm{NN}}$ = 5.02 TeV}},
                    }{}\href{https://doi.org/10.1016/j.physletb.2022.137397}{Phys.\ Lett.\
                    \textbf{B835} (2022) 137397},
                    \href{http://arxiv.org/abs/2202.11807}{{\normalfont\ttfamily
                    arXiv:2202.11807}}\relax
                    \mciteBstWouldAddEndPuncttrue
                    \mciteSetBstMidEndSepPunct{\mcitedefaultmidpunct}
                    {\mcitedefaultendpunct}{\mcitedefaultseppunct}\relax
                    \EndOfBibitem
                \bibitem{CMS:2017exb}
                    CMS collaboration, A.~M. Sirunyan {\em et~al.},
                    \ifthenelse{\boolean{articletitles}}{\emph{{Measurement of prompt and
                    nonprompt \jpsi production in pp and pPb collisions at $\sqsnn = $5.02\tev}},
                    }{}\href{https://doi.org/10.1140/epjc/s10052-017-4828-3}{Eur.\ Phys.\ J.\
                    \textbf{C77} (2017) 269},
                    \href{http://arxiv.org/abs/1702.01462}{{\normalfont\ttfamily
                    arXiv:1702.01462}}\relax
                    \mciteBstWouldAddEndPuncttrue
                    \mciteSetBstMidEndSepPunct{\mcitedefaultmidpunct}
                    {\mcitedefaultendpunct}{\mcitedefaultseppunct}\relax
                    \EndOfBibitem
                \bibitem{CMS:2015sfx}
                    CMS collaboration, V.~Khachatryan {\em et~al.},
                    \ifthenelse{\boolean{articletitles}}{\emph{{Study of B meson production in
                    p+Pb collisions at $\sqsnn = $5.02\tev using exclusive hadronic decays}},
                    }{}\href{https://doi.org/10.1103/PhysRevLett.116.032301}{Phys.\ Rev.\ Lett.\
                    \textbf{116} (2016) 032301},
                    \href{http://arxiv.org/abs/1508.06678}{{\normalfont\ttfamily
                    arXiv:1508.06678}}\relax
                    \mciteBstWouldAddEndPuncttrue
                    \mciteSetBstMidEndSepPunct{\mcitedefaultmidpunct}
                    {\mcitedefaultendpunct}{\mcitedefaultseppunct}\relax
                    \EndOfBibitem
                \bibitem{CMS:2016wma}
                    CMS collaboration, A.~M. Sirunyan {\em et~al.},
                    \ifthenelse{\boolean{articletitles}}{\emph{{Measurements of the charm jet
                    cross section and nuclear modification factor in pPb collisions at $\sqsnn$ =
                    5.02\tev}}, }{}\href{https://doi.org/10.1016/j.physletb.2017.06.053}{Phys.\
                    Lett.\ \textbf{B772} (2017) 306},
                    \href{http://arxiv.org/abs/1612.08972}{{\normalfont\ttfamily
                    arXiv:1612.08972}}\relax
                    \mciteBstWouldAddEndPuncttrue
                    \mciteSetBstMidEndSepPunct{\mcitedefaultmidpunct}
                    {\mcitedefaultendpunct}{\mcitedefaultseppunct}\relax
                    \EndOfBibitem
                \bibitem{CMS:2015gcq}
                    CMS collaboration, V.~Khachatryan {\em et~al.},
                    \ifthenelse{\boolean{articletitles}}{\emph{{Transverse momentum spectra of
                    inclusive b jets in pPb collisions at $\sqsnn = $5.02\tev}},
                    }{}\href{https://doi.org/10.1016/j.physletb.2016.01.010}{Phys.\ Lett.\
                    \textbf{B754} (2016) 59},
                    \href{http://arxiv.org/abs/1510.03373}{{\normalfont\ttfamily
                    arXiv:1510.03373}}\relax
                    \mciteBstWouldAddEndPuncttrue
                    \mciteSetBstMidEndSepPunct{\mcitedefaultmidpunct}
                    {\mcitedefaultendpunct}{\mcitedefaultseppunct}\relax
                    \EndOfBibitem
                \bibitem{STAR:2004ocv}
                    STAR collaboration, J.~Adams {\em et~al.},
                    \ifthenelse{\boolean{articletitles}}{\emph{{Open charm yields in d + Au
                    collisions at $\sqsnn = $200\gev}},
                    }{}\href{https://doi.org/10.1103/PhysRevLett.94.062301}{Phys.\ Rev.\ Lett.\
                    \textbf{94} (2005) 062301},
                    \href{http://arxiv.org/abs/nucl-ex/0407006}{{\normalfont\ttfamily
                    arXiv:nucl-ex/0407006}}\relax
                    \mciteBstWouldAddEndPuncttrue
                    \mciteSetBstMidEndSepPunct{\mcitedefaultmidpunct}
                    {\mcitedefaultendpunct}{\mcitedefaultseppunct}\relax
                    \EndOfBibitem
                \bibitem{PHENIX:2012hww}
                    PHENIX collaboration, A.~Adare {\em et~al.},
                    \ifthenelse{\boolean{articletitles}}{\emph{{Cold-nuclear-matter effects on
                    heavy-quark production in $d+$Au collisions at $\sqsnn = $200\gev}},
                    }{}\href{https://doi.org/10.1103/PhysRevLett.109.242301}{Phys.\ Rev.\ Lett.\
                    \textbf{109} (2012) 242301},
                    \href{http://arxiv.org/abs/1208.1293}{{\normalfont\ttfamily
                    arXiv:1208.1293}}\relax
                    \mciteBstWouldAddEndPuncttrue
                    \mciteSetBstMidEndSepPunct{\mcitedefaultmidpunct}
                    {\mcitedefaultendpunct}{\mcitedefaultseppunct}\relax
                    \EndOfBibitem
                \bibitem{Eskola:2019bgf}
                    K.~J. Eskola, I.~Helenius, P.~Paakkinen, and H.~Paukkunen,
                    \ifthenelse{\boolean{articletitles}}{\emph{{A QCD analysis of LHCb D-meson
                    data in p+Pb collisions}},
                    }{}\href{https://doi.org/10.1007/JHEP05(2020)037}{JHEP \textbf{05} (2020)
                    037}, \href{http://arxiv.org/abs/1906.02512}{{\normalfont\ttfamily
                    arXiv:1906.02512}}\relax
                    \mciteBstWouldAddEndPuncttrue
                    \mciteSetBstMidEndSepPunct{\mcitedefaultmidpunct}
                    {\mcitedefaultendpunct}{\mcitedefaultseppunct}\relax
                    \EndOfBibitem
                \bibitem{Khalek:2022zqe}
                    R.~A. Khalek {\em et~al.},
                    \ifthenelse{\boolean{articletitles}}{\emph{{nNNPDF3.0: Evidence for a
                    modified partonic structure in heavy nuclei}},
                    }{}\href{http://arxiv.org/abs/2201.12363}{{\normalfont\ttfamily
                    arXiv:2201.12363}}\relax
                    \mciteBstWouldAddEndPuncttrue
                    \mciteSetBstMidEndSepPunct{\mcitedefaultmidpunct}
                    {\mcitedefaultendpunct}{\mcitedefaultseppunct}\relax
                    \EndOfBibitem
                \bibitem{Alves:2008zz}
                    LHCb collaboration, A.~A. Alves~Jr.\ {\em et~al.},
                    \ifthenelse{\boolean{articletitles}}{\emph{{The \lhcb detector at the LHC}},
                    }{}\href{https://doi.org/10.1088/1748-0221/3/08/S08005}{JINST \textbf{3}
                    (2008) S08005}\relax
                    \mciteBstWouldAddEndPuncttrue
                    \mciteSetBstMidEndSepPunct{\mcitedefaultmidpunct}
                    {\mcitedefaultendpunct}{\mcitedefaultseppunct}\relax
                    \EndOfBibitem
                \bibitem{LHCb-DP-2014-002}
                    LHCb collaboration, R.~Aaij {\em et~al.},
                    \ifthenelse{\boolean{articletitles}}{\emph{{LHCb detector performance}},
                    }{}\href{https://doi.org/10.1142/S0217751X15300227}{Int.\ J.\ Mod.\ Phys.\
                    \textbf{A30} (2015) 1530022},
                    \href{http://arxiv.org/abs/1412.6352}{{\normalfont\ttfamily
                    arXiv:1412.6352}}\relax
                    \mciteBstWouldAddEndPuncttrue
                    \mciteSetBstMidEndSepPunct{\mcitedefaultmidpunct}
                    {\mcitedefaultendpunct}{\mcitedefaultseppunct}\relax
                    \EndOfBibitem
                \bibitem{LHCb-PAPER-2014-047}
                    LHCb collaboration, R.~Aaij {\em et~al.},
                    \ifthenelse{\boolean{articletitles}}{\emph{{Precision luminosity measurements
                    at LHCb}}, }{}\href{https://doi.org/10.1088/1748-0221/9/12/P12005}{JINST
                    \textbf{9} (2014) P12005},
                    \href{http://arxiv.org/abs/1410.0149}{{\normalfont\ttfamily
                    arXiv:1410.0149}}\relax
                    \mciteBstWouldAddEndPuncttrue
                    \mciteSetBstMidEndSepPunct{\mcitedefaultmidpunct}
                    {\mcitedefaultendpunct}{\mcitedefaultseppunct}\relax
                    \EndOfBibitem
                \bibitem{Sjostrand:2006za}
                    T.~Sj\"{o}strand, S.~Mrenna, and P.~Skands,
                    \ifthenelse{\boolean{articletitles}}{\emph{{PYTHIA 6.4 physics and manual}},
                    }{}\href{https://doi.org/10.1088/1126-6708/2006/05/026}{JHEP \textbf{05}
                    (2006) 026}, \href{http://arxiv.org/abs/hep-ph/0603175}{{\normalfont\ttfamily
                    arXiv:hep-ph/0603175}}\relax
                    \mciteBstWouldAddEndPuncttrue
                    \mciteSetBstMidEndSepPunct{\mcitedefaultmidpunct}
                    {\mcitedefaultendpunct}{\mcitedefaultseppunct}\relax
                    \EndOfBibitem
                \bibitem{Sjostrand:2007gs}
                    T.~Sj\"{o}strand, S.~Mrenna, and P.~Skands,
                    \ifthenelse{\boolean{articletitles}}{\emph{{A brief introduction to PYTHIA
                    8.1}}, }{}\href{https://doi.org/10.1016/j.cpc.2008.01.036}{Comput.\ Phys.\
                    Commun.\  \textbf{178} (2008) 852},
                    \href{http://arxiv.org/abs/0710.3820}{{\normalfont\ttfamily
                    arXiv:0710.3820}}\relax
                    \mciteBstWouldAddEndPuncttrue
                    \mciteSetBstMidEndSepPunct{\mcitedefaultmidpunct}
                    {\mcitedefaultendpunct}{\mcitedefaultseppunct}\relax
                    \EndOfBibitem
                \bibitem{PhysRevC.92.034906}
                    T.~Pierog {\em et~al.}, \ifthenelse{\boolean{articletitles}}{\emph{{EPOS}
                    {LHC}: Test of collective hadronization with data measured at the {CERN}
                    {Large Hadron Collider}},
                    }{}\href{https://doi.org/10.1103/PhysRevC.92.034906}{Phys.\ Rev.\
                    \textbf{C92} (2015) 034906}\relax
                    \mciteBstWouldAddEndPuncttrue
                    \mciteSetBstMidEndSepPunct{\mcitedefaultmidpunct}
                    {\mcitedefaultendpunct}{\mcitedefaultseppunct}\relax
                    \EndOfBibitem
                \bibitem{LHCb-PROC-2010-056}
                    I.~Belyaev {\em et~al.}, \ifthenelse{\boolean{articletitles}}{\emph{{Handling
                    of the generation of primary events in Gauss, the LHCb simulation
                    framework}}, }{}\href{https://doi.org/10.1088/1742-6596/331/3/032047}{J.\
                    Phys.\ Conf.\ Ser.\  \textbf{331} (2011) 032047}\relax
                    \mciteBstWouldAddEndPuncttrue
                    \mciteSetBstMidEndSepPunct{\mcitedefaultmidpunct}
                    {\mcitedefaultendpunct}{\mcitedefaultseppunct}\relax
                    \EndOfBibitem
                \bibitem{Lange:2001uf}
                    D.~J. Lange, \ifthenelse{\boolean{articletitles}}{\emph{{The EvtGen particle
                    decay simulation package}},
                    }{}\href{https://doi.org/10.1016/S0168-9002(01)00089-4}{Nucl.\ Instrum.\
                    Meth.\  \textbf{A462} (2001) 152}\relax
                    \mciteBstWouldAddEndPuncttrue
                    \mciteSetBstMidEndSepPunct{\mcitedefaultmidpunct}
                    {\mcitedefaultendpunct}{\mcitedefaultseppunct}\relax
                    \EndOfBibitem
                \bibitem{Golonka:2005pn}
                    P.~Golonka and Z.~Was, \ifthenelse{\boolean{articletitles}}{\emph{{PHOTOS Monte
                    Carlo: A precision tool for QED corrections in $Z$ and $W$ decays}},
                    }{}\href{https://doi.org/10.1140/epjc/s2005-02396-4}{Eur.\ Phys.\ J.\
                    \textbf{C45} (2006) 97},
                    \href{http://arxiv.org/abs/hep-ph/0506026}{{\normalfont\ttfamily
                    arXiv:hep-ph/0506026}}\relax
                    \mciteBstWouldAddEndPuncttrue
                    \mciteSetBstMidEndSepPunct{\mcitedefaultmidpunct}
                    {\mcitedefaultendpunct}{\mcitedefaultseppunct}\relax
                    \EndOfBibitem
                \bibitem{Allison:2006ve}
                    Geant4 collaboration, J.~Allison {\em et~al.},
                    \ifthenelse{\boolean{articletitles}}{\emph{{Geant4 developments and
                    applications}}, }{}\href{https://doi.org/10.1109/TNS.2006.869826}{IEEE
                    Trans.\ Nucl.\ Sci.\  \textbf{53} (2006) 270}\relax
                    \mciteBstWouldAddEndPuncttrue
                    \mciteSetBstMidEndSepPunct{\mcitedefaultmidpunct}
                    {\mcitedefaultendpunct}{\mcitedefaultseppunct}\relax
                    \EndOfBibitem
                \bibitem{Agostinelli:2002hh}
                    Geant4 collaboration, S.~Agostinelli {\em et~al.},
                    \ifthenelse{\boolean{articletitles}}{\emph{{Geant4: A simulation toolkit}},
                    }{}\href{https://doi.org/10.1016/S0168-9002(03)01368-8}{Nucl.\ Instrum.\
                    Meth.\  \textbf{A506} (2003) 250}\relax
                    \mciteBstWouldAddEndPuncttrue
                    \mciteSetBstMidEndSepPunct{\mcitedefaultmidpunct}
                    {\mcitedefaultendpunct}{\mcitedefaultseppunct}\relax
                    \EndOfBibitem
                \bibitem{LHCb-PROC-2011-006}
                    M.~Clemencic {\em et~al.}, \ifthenelse{\boolean{articletitles}}{\emph{{The
                    \lhcb simulation application, Gauss: Design, evolution and experience}},
                    }{}\href{https://doi.org/10.1088/1742-6596/331/3/032023}{J.\ Phys.\ Conf.\
                    Ser.\  \textbf{331} (2011) 032023}\relax
                    \mciteBstWouldAddEndPuncttrue
                    \mciteSetBstMidEndSepPunct{\mcitedefaultmidpunct}
                    {\mcitedefaultendpunct}{\mcitedefaultseppunct}\relax
                    \EndOfBibitem
                \bibitem{PDG2022}
                    Particle Data Group, R.~L. Workman {\em et~al.},
                    \ifthenelse{\boolean{articletitles}}{\emph{{Review of Particle Physics}},
                    }{}\href{https://doi.org/10.1093/ptep/ptac097}{PTEP \textbf{2022} (2022)
                    083C01}\relax
                    \mciteBstWouldAddEndPuncttrue
                    \mciteSetBstMidEndSepPunct{\mcitedefaultmidpunct}
                    {\mcitedefaultendpunct}{\mcitedefaultseppunct}\relax
                    \EndOfBibitem
                \bibitem{Skwarnicki:1986xj}
                    T.~Skwarnicki, {\em {A study of the radiative cascade transitions between the
                    Upsilon-prime and Upsilon resonances}}, PhD thesis, Institute of Nuclear
                    Physics, Krakow, 1986,
                    {\href{http://inspirehep.net/record/230779/}{DESY-F31-86-02}}\relax
                    \mciteBstWouldAddEndPuncttrue
                    \mciteSetBstMidEndSepPunct{\mcitedefaultmidpunct}
                    {\mcitedefaultendpunct}{\mcitedefaultseppunct}\relax
                    \EndOfBibitem
                \bibitem{pivk2005plots}
                    M.~Pivk and F.~R. Le~Diberder,
                    \ifthenelse{\boolean{articletitles}}{\emph{{sPlot: A statistical tool to
                    unfold data distributions}},
                    }{}\href{https://doi.org/10.1016/j.nima.2005.08.106}{Nucl.\ Instrum.\ Meth.\
                    \textbf{A555} (2005) 356},
                    \href{http://arxiv.org/abs/physics/0402083}{{\normalfont\ttfamily
                    arXiv:physics/0402083}}\relax
                    \mciteBstWouldAddEndPuncttrue
                    \mciteSetBstMidEndSepPunct{\mcitedefaultmidpunct}
                    {\mcitedefaultendpunct}{\mcitedefaultseppunct}\relax
                    \EndOfBibitem
                \bibitem{bukin2007fitting}
                    A.~D. Bukin, \ifthenelse{\boolean{articletitles}}{\emph{Fitting function for
                    asymmetric peaks},
                    }{}\href{http://arxiv.org/abs/0711.4449}{{\normalfont\ttfamily
                    arXiv:0711.4449}}\relax
                    \mciteBstWouldAddEndPuncttrue
                    \mciteSetBstMidEndSepPunct{\mcitedefaultmidpunct}
                    {\mcitedefaultendpunct}{\mcitedefaultseppunct}\relax
                    \EndOfBibitem
                \bibitem{ref:SuppMat}                         
                    {See Supplemental Material at [link inserted by publisher] for a summary of
                    systematic uncertainties and numerical results and additional plots for the 
                    fit result, nuclear modification factor and forward-backward production
                    ratio}\relax
                    \mciteBstWouldAddEndPuncttrue
                    \mciteSetBstMidEndSepPunct{\mcitedefaultmidpunct}
                    {\mcitedefaultendpunct}{\mcitedefaultseppunct}\relax
                    \EndOfBibitem
                \bibitem{LHCb-DP-2013-002}
                    LHCb collaboration, R.~Aaij {\em et~al.},
                    \ifthenelse{\boolean{articletitles}}{\emph{{Measurement of the track
                    reconstruction efficiency at LHCb}},
                    }{}\href{https://doi.org/10.1088/1748-0221/10/02/P02007}{JINST \textbf{10}
                    (2015) P02007}, \href{http://arxiv.org/abs/1408.1251}{{\normalfont\ttfamily
                    arXiv:1408.1251}}\relax
                    \mciteBstWouldAddEndPuncttrue
                    \mciteSetBstMidEndSepPunct{\mcitedefaultmidpunct}
                    {\mcitedefaultendpunct}{\mcitedefaultseppunct}\relax
                    \EndOfBibitem
                \bibitem{LHCb-PUB-2016-021}
                    L.~Anderlini {\em et~al.}, \ifthenelse{\boolean{articletitles}}{\emph{{The
                    PIDCalib package}}, }{}
                    \href{http://cdsweb.cern.ch/search?p=LHCb-PUB-2016-021&f=reportnumber&action_search=Search&c=LHCb+Notes}
                    {LHCb-PUB-2016-021}, 2016\relax
                    \mciteBstWouldAddEndPuncttrue
                    \mciteSetBstMidEndSepPunct{\mcitedefaultmidpunct}
                    {\mcitedefaultendpunct}{\mcitedefaultseppunct}\relax
                    \EndOfBibitem
                \bibitem{LHCb-DP-2018-001}
                    R.~Aaij {\em et~al.}, \ifthenelse{\boolean{articletitles}}{\emph{{Selection and
                    processing of calibration samples to measure the particle identification
                    performance of the LHCb experiment in Run 2}},
                    }{}\href{https://doi.org/10.1140/epjti/s40485-019-0050-z}{Eur.\ Phys.\ J.\
                    Tech.\ Instr.\  \textbf{6} (2019) 1},
                    \href{http://arxiv.org/abs/1803.00824}{{\normalfont\ttfamily
                    arXiv:1803.00824}}\relax
                    \mciteBstWouldAddEndPuncttrue
                    \mciteSetBstMidEndSepPunct{\mcitedefaultmidpunct}
                    {\mcitedefaultendpunct}{\mcitedefaultseppunct}\relax
                    \EndOfBibitem
                \bibitem{LHCb-PAPER-2016-042}
                    LHCb collaboration, R.~Aaij {\em et~al.},
                    \ifthenelse{\boolean{articletitles}}{\emph{{Measurements of prompt charm
                    production cross-sections in \proton\proton collisions at $\sqs = $5\tev}},
                    }{}\href{https://doi.org/10.1007/JHEP06(2017)147}{JHEP \textbf{06} (2017)
                    147}, \href{http://arxiv.org/abs/1610.02230}{{\normalfont\ttfamily
                    arXiv:1610.02230}}\relax
                    \mciteBstWouldAddEndPuncttrue
                    \mciteSetBstMidEndSepPunct{\mcitedefaultmidpunct}
                    {\mcitedefaultendpunct}{\mcitedefaultseppunct}\relax
                    \EndOfBibitem
                \bibitem{LHCb-PAPER-2015-041}
                    LHCb collaboration, R.~Aaij {\em et~al.},
                    \ifthenelse{\boolean{articletitles}}{\emph{{Measurements of prompt charm
                    production cross-sections in \proton\proton collisions at $\sqs = $13\tev}},
                    }{}\href{https://doi.org/10.1007/JHEP03(2016)159}{JHEP \textbf{03} (2016)
                    159}, Erratum \href{https://doi.org/10.1007/JHEP09(2016)013}{ibid.\
                    \textbf{09} (2016) 013}, Erratum
                    \href{https://doi.org/10.1007/JHEP05(2017)074}{ibid.\   \textbf{05} (2017)
                    074}, \href{http://arxiv.org/abs/1510.01707}{{\normalfont\ttfamily
                    arXiv:1510.01707}}\relax
                    \mciteBstWouldAddEndPuncttrue
                    \mciteSetBstMidEndSepPunct{\mcitedefaultmidpunct}
                    {\mcitedefaultendpunct}{\mcitedefaultseppunct}\relax
                    \EndOfBibitem
                \bibitem{PhysRevLett.121.052004}
                    A.~Kusina, J.-P. Lansberg, I.~Schienbein, and H.-S. Shao,
                    \ifthenelse{\boolean{articletitles}}{\emph{Gluon shadowing in heavy-flavor
                    production at the lhc},
                    }{}\href{https://doi.org/10.1103/PhysRevLett.121.052004}{Phys.\ Rev.\ Lett.\
                    \textbf{121} (2018) 052004}\relax
                    \mciteBstWouldAddEndPuncttrue
                    \mciteSetBstMidEndSepPunct{\mcitedefaultmidpunct}
                    {\mcitedefaultendpunct}{\mcitedefaultseppunct}\relax
                    \EndOfBibitem
                \bibitem{Shao:2012iz}
                    H.-S. Shao, \ifthenelse{\boolean{articletitles}}{\emph{{HELAC-Onia: An
                    automatic matrix element generator for heavy quarkonium physics}},
                    }{}\href{https://doi.org/10.1016/j.cpc.2013.05.023}{Comput.\ Phys.\ Commun.\
                    \textbf{184} (2013) 2562},
                    \href{http://arxiv.org/abs/1212.5293}{{\normalfont\ttfamily
                    arXiv:1212.5293}}\relax
                    \mciteBstWouldAddEndPuncttrue
                    \mciteSetBstMidEndSepPunct{\mcitedefaultmidpunct}
                    {\mcitedefaultendpunct}{\mcitedefaultseppunct}\relax
                    \EndOfBibitem
                \bibitem{Shao:2015vga}
                    H.-S. Shao, \ifthenelse{\boolean{articletitles}}{\emph{{HELAC-Onia 2.0: an
                    upgraded matrix-element and event generator for heavy quarkonium physics}},
                    }{}\href{https://doi.org/10.1016/j.cpc.2015.09.011}{Comput.\ Phys.\ Commun.\
                    \textbf{198} (2016) 238},
                    \href{http://arxiv.org/abs/1507.03435}{{\normalfont\ttfamily
                    arXiv:1507.03435}}\relax
                    \mciteBstWouldAddEndPuncttrue
                    \mciteSetBstMidEndSepPunct{\mcitedefaultmidpunct}
                    {\mcitedefaultendpunct}{\mcitedefaultseppunct}\relax
                    \EndOfBibitem
                \bibitem{Eskola:2016oht}
                    K.~J. Eskola, P.~Paakkinen, H.~Paukkunen, and C.~A. Salgado,
                    \ifthenelse{\boolean{articletitles}}{\emph{{EPPS16: Nuclear parton
                    distributions with LHC data}},
                    }{}\href{https://doi.org/10.1140/epjc/s10052-017-4725-9}{Eur.\ Phys.\ J.\
                    \textbf{C77} (2017) 163},
                    \href{http://arxiv.org/abs/1612.05741}{{\normalfont\ttfamily
                    arXiv:1612.05741}}\relax
                    \mciteBstWouldAddEndPuncttrue
                    \mciteSetBstMidEndSepPunct{\mcitedefaultmidpunct}
                    {\mcitedefaultendpunct}{\mcitedefaultseppunct}\relax
                    \EndOfBibitem
                \bibitem{Kovarik:2015cma}
                    K.~Kovarik {\em et~al.}, \ifthenelse{\boolean{articletitles}}{\emph{{nCTEQ15 -
                    Global analysis of nuclear parton distributions with uncertainties in the
                    CTEQ framework}}, }{}\href{https://doi.org/10.1103/PhysRevD.93.085037}{Phys.\
                    Rev.\  \textbf{D93} (2016) 085037},
                    \href{http://arxiv.org/abs/1509.00792}{{\normalfont\ttfamily
                    arXiv:1509.00792}}\relax
                    \mciteBstWouldAddEndPuncttrue
                    \mciteSetBstMidEndSepPunct{\mcitedefaultmidpunct}
                    {\mcitedefaultendpunct}{\mcitedefaultseppunct}\relax
                    \EndOfBibitem
                \bibitem{Lansberg:2016deg}
                    J.-P. Lansberg and H.-S. Shao,
                    \ifthenelse{\boolean{articletitles}}{\emph{{Towards an automated tool to
                    evaluate the impact of the nuclear modification of the gluon density on
                    quarkonium, D and B meson production in proton–nucleus collisions}},
                    }{}\href{https://doi.org/10.1140/epjc/s10052-016-4575-x}{Eur.\ Phys.\ J.\
                    \textbf{C77} (2017) 1},
                    \href{http://arxiv.org/abs/1610.05382}{{\normalfont\ttfamily
                    arXiv:1610.05382}}\relax
                    \mciteBstWouldAddEndPuncttrue
                    \mciteSetBstMidEndSepPunct{\mcitedefaultmidpunct}
                    {\mcitedefaultendpunct}{\mcitedefaultseppunct}\relax
                    \EndOfBibitem
                \bibitem{Ducloue:2015gfa}
                    B.~Duclou\'e, T.~Lappi, and H.~M\"antysaari,
                    \ifthenelse{\boolean{articletitles}}{\emph{{Forward $J/\psi$ production in
                    proton-nucleus collisions at high energy}},
                    }{}\href{https://doi.org/10.1103/PhysRevD.91.114005}{Phys.\ Rev.\
                    \textbf{D91} (2015) 114005},
                    \href{http://arxiv.org/abs/1503.02789}{{\normalfont\ttfamily
                    arXiv:1503.02789}}\relax
                    \mciteBstWouldAddEndPuncttrue
                    \mciteSetBstMidEndSepPunct{\mcitedefaultmidpunct}
                    {\mcitedefaultendpunct}{\mcitedefaultseppunct}\relax
                    \EndOfBibitem
                \bibitem{Ducloue:2016ywt}
                    B.~Duclou\'e, T.~Lappi, and H.~M\"antysaari,
                    \ifthenelse{\boolean{articletitles}}{\emph{{Forward $J/\psi$ and $D$ meson
                    nuclear suppression at the LHC}},
                    }{}\href{https://doi.org/10.1016/j.nuclphysbps.2017.05.071}{Nucl.\ Part.\
                    Phys.\ Proc.\  \textbf{289-290} (2017) 309},
                    \href{http://arxiv.org/abs/1612.04585}{{\normalfont\ttfamily
                    arXiv:1612.04585}}\relax
                    \mciteBstWouldAddEndPuncttrue
                    \mciteSetBstMidEndSepPunct{\mcitedefaultmidpunct}
                    {\mcitedefaultendpunct}{\mcitedefaultseppunct}\relax
                    \EndOfBibitem
                \bibitem{Ma:2018bax}
                    Y.-Q. Ma, P.~Tribedy, R.~Venugopalan, and K.~Watanabe,
                    \ifthenelse{\boolean{articletitles}}{\emph{{Event engineering studies for
                    heavy flavor production and hadronization in high multiplicity hadron-hadron
                    and hadron-nucleus collisions}},
                    }{}\href{https://doi.org/10.1103/PhysRevD.98.074025}{Phys.\ Rev.\
                    \textbf{D98} (2018) 074025},
                    \href{http://arxiv.org/abs/1803.11093}{{\normalfont\ttfamily
                    arXiv:1803.11093}}\relax
                    \mciteBstWouldAddEndPuncttrue
                    \mciteSetBstMidEndSepPunct{\mcitedefaultmidpunct}
                    {\mcitedefaultendpunct}{\mcitedefaultseppunct}\relax
                    \EndOfBibitem
                \bibitem{PhysRevLett.100.022303}
                    H. Kowalski, T. Lappi, and R. Venugopalan, 
                    \ifthenelse{\boolean{articletitles}}{\emph{Nuclear enhancement of universal dynamics of 
                    high parton densities},
                    }{}\href{https://doi.org/10.1103/PhysRevLett.100.022303}{Phys.\ Rev.\ Lett.\
                    \textbf{100} (2008) 022303}
                    \href{http://arxiv.org/abs/0705.3047}{{\normalfont\ttfamily
                    arXiv:0705.3047}}\relax
                    \mciteBstWouldAddEndPuncttrue
                    \mciteSetBstMidEndSepPunct{\mcitedefaultmidpunct}
                    {\mcitedefaultendpunct}{\mcitedefaultseppunct}\relax
                    \EndOfBibitem
            \end{mcitethebibliography}
